\documentclass[10pt]{article}

\usepackage{amsmath}

\usepackage{array}

\usepackage{appendix}

\usepackage{tocloft}                   

\usepackage{graphicx}

\usepackage{amsfonts}

\usepackage{amssymb}

\usepackage{mathrsfs}

\usepackage{yfonts}

\usepackage{euscript}

\usepackage{centernot}                 

\usepackage{ifsym}                     

\usepackage{upgreek}

\usepackage{mathtools}

\usepackage{color}

\usepackage{slantsc}
\usepackage{calligra}

\usepackage{bbold}          

\usepackage[T1]{fontenc}

\usepackage{epsf}

\usepackage{latexsym}

\usepackage{tipa}

\usepackage{makeidx}

\makeindex

\def\be{\begin{equation}}

\def\ee{\end{equation}}

\def\beq{\begin{equation}}

\def\eeq{\end{equation}}

\def\bea{\begin{eqnarray}}

\def\eea{\end{eqnarray}}

\def\ni{\noindent}

\def\!{\hspace{-1.6667em}}

\def\slTheta{\mathit{\Theta}}                     

\def\slPhi{\mathit{\Phi}}                         

\def\bigiota{\mbox{\Large $\iota$}}                 

\def\bigsigma{\mbox{\Large $\sigma$}}

\def\uiR{\underline{R}}

\def\uiq{\underline{q}}

\def\uir{\underline{r}}

\def\mF{\mbox{F}}

\def\mT{\mbox{T}}

\def\uR{\underline{\mbox{$R$}}}

\def\uq{\underline{\mbox{q}}} 

\def\ur{\underline{\mbox{r}}}

\def\urho{{\underline{\rho}}}

\def\sa{\mbox{\scriptsize a}}

\def\si{\mbox{\scriptsize i}}

\def\sm{\mbox{\scriptsize m}}

\def\sn{\mbox{\scriptsize n}}

\def\sss{\mbox{\scriptsize s}}  

\def\sx{\mbox{\scriptsize x}}

\def\sA{\mbox{\scriptsize A}}

\def\sG{\mbox{\scriptsize G}}

\def\sM{\mbox{\scriptsize M}}

\def\sQ{\mbox{\scriptsize Q}} 

\def\sR{\mbox{\scriptsize R}}

\def\sS{\mbox{\scriptsize S}}

\def\sT{\mbox{\scriptsize T}}

\def\bigupalpha{\mbox{\Large$\alpha$}}

\def\bigmu{\mbox{\Large$\mu$}}

\def\cr{\mbox{\scriptsize{\bf $\mbox{ } \times \mbox{ }$}}}

\def\sumi2{\sum\mbox{}_{\mbox{}_{\mbox{\scriptsize $i$=1}}}^2}

\def\sumi3{\sum\mbox{}_{\mbox{}_{\mbox{\scriptsize $i$=1}}}^3}

\def\sumABcycles3{\sum\mbox{}_{\mbox{}_{\mbox{\scriptsize cycles $A,B$=1}}}^{3}}

\def\sumCDcycles3{\sum\mbox{}_{\mbox{}_{\mbox{\scriptsize cycles $C,D$=1}}}^{3}}

\def\sumj3{\sum\mbox{}_{\mbox{}_{\mbox{\scriptsize $j$=1}}}^3}

\def\sumk3{\sum\mbox{}_{\mbox{}_{\mbox{\scriptsize $k$=1}}}^3}

\def\prodiA1{\prod\mbox{}_{\mbox{}_{\mbox{\scriptsize $i$=1}}}^{A - 1}}

\def\d{\textrm{d}}                                                  

\def\FrS{\mbox{\Large $\mathfrak{s}$}}                         

\def\Hilb{\mbox{{\boldmath$\mathfrak{H}$}ilb}}                 

\def\scN{\mbox{\scriptsize ${\cal N}$}}

\def\Phase{\mbox{{\boldmath$\mathfrak{P}$}hase}}                     

\def\bFrR{\mbox{\boldmath$\mathfrak{R}$}}                            
\def\Rig-Phase{\bFrR\mbox{ig-}\Phase}                                
                                                                													   
\def\FrP{\mbox{\Large $\mathfrak{p}$}}                                 
\def\Positive-Modespace{\mbox{{\boldmath$\mathfrak{M}$}odespace$^+$}}

\def\POSITIVE-MODESPACE{\mbox{{\boldmath$\mathfrak{M}$}ODESPACE$^+$}}

\def\Kin-Hilb{\mbox{{\boldmath$\mathfrak{K}$}in-\Hilb}}                     

\def\Mid-Hilb{\mbox{{\boldmath$\mathfrak{M}$}id-\Hilb}}                     

\def\Dyn-Hilb{\mbox{{\boldmath$\mathfrak{D}$}yn-\Hilb}}                     

\def\5Star{\mbox{\Large$\star$}}              

\def\Frr{\mbox{$\mathfrak{r}$}}

\begin{document}

\begin{titlepage}

\begin{center}

{\bf \Large Shape (In)dependent Inequalities}

\mbox{ }

{\bf for Triangleland's Jacobi and Democratic-Linear Ellipticity Quantitities}

\mbox{ }

{\large \bf Edward Anderson$^*$}

\vspace{.2in}

\end{center}

\begin{abstract}

Sides and medians are both Jacobi coordinate magnitudes, moreover then equably entering the spherical coordinates on Kendall's shape sphere and the Hopf coordinates. 
This motivates treating medians on the same footing as sides in triangle geometry and the resulting Shape Theory. 
In this paper, we consequently reformulate inequalities for the medians in terms of shape quantities, and proceed to find inequalities on the mass-weighted Jacobi coordinates.
This work moreover identifies the $4/3$ -- powers of which occur frequently in the theory of medians -- as the ratio of Jacobi masses.

\mbox{ }

\ni One of the Hopf coordinates is tetra-area. 
Another is anisoscelesness, which parametrizes whether triangles are left-or-right leaning as bounded by isoscelesness itself.   
The third is ellipticity, which parametrizes tallness-or-flatness of triangles as bounded by regular triangles. 
Whereas tetra-area is clearly cluster choice invariant, 
Jacobi coordinates, anisoscelesness and ellipticity are cluster choice dependent but can be `democratized' by averaging over all clusters. 
Democratized ellipticity moreover trivializes, due to ellipticity being the difference of base-side and median second moments, whose averages are equal to each other.  
Thus we introduce a distinct `linear ellipticity' quantifier of tallness-or-flatness of triangles whose democratization is nontrivial, and find inequalities bounding this. 
Some of this paper's inequalities are shape-independent bounds, whereas others' bounds depend on the isoperimetric ratio and arithmetic-to-geometric side mean ratio shape variables. 

\end{abstract}

\mbox{ }

\ni Keywords:             Applied Geometry, 
                          Background Independence, 
                          Shape Theory, 
                          Relationalism, 						  
                          triangles  
						  Kendall's Shape Statistics, 
						  geometrical inequalities, 
						  relative Jacobi coordinates, 
						  Hopf fibration,
						  3-body problem.

\vspace{0.1in}
  
\ni $^*$ Dr.E.Anderson.Maths.Physics@protonmail.com

\end{titlepage}

\section{Introduction}
%
{            \begin{figure}[!ht]
\centering
\includegraphics[width=0.25\textwidth]{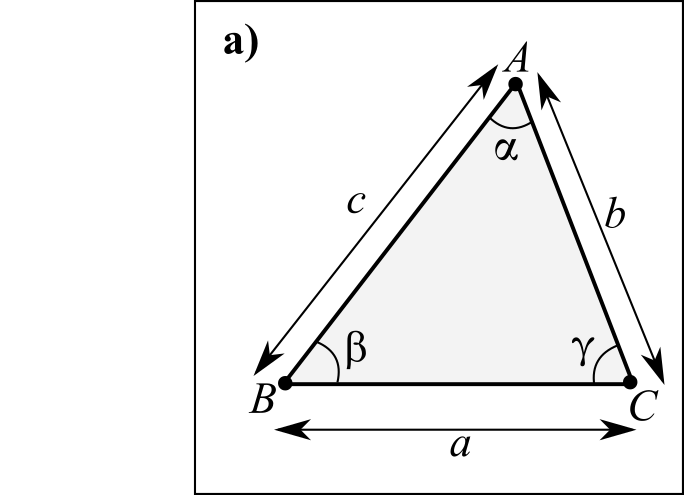}
\caption[Text der im Bilderverzeichnis auftaucht]{        \footnotesize{a) Labelling of vertices and edges of the triangle.  }  } 
\label{Triangle-Notation} \end{figure}           }

\ni For the arbitrary triangle -- $\triangle\,ABC$ with vertices A, B, C -- side-lengths and angles between sides are usually considered to be primary information.
In the current paper, we denote these in the customary cyclic manner of Fig \ref{Triangle-Notation}.  
Denoting $a$, $b$, $c$ by $s_i$, $i = 1$ to $3$ will also be useful for us.   

\mbox{ }

\ni Many of the current paper's innovations start moreover by taking due note of the $N$-body problem formulation of the triangle viewed as a 3-body problem, 
with point-or-particle positions corresponding to the vertices A, B, C. 
Position coordinates for these are given in Fig \ref{Jac-Med-Ineq-Fig-2}.a).
These are with respect to an absolute origin 0, absolute axes $A$ and absolute scale $S$.
One can moreover strip away these absolute 
(in Physics, or `carrier space' \cite{I} in a wider context that pins no physical space significance on the flat space the triangle in question is realized in) features. 
The first parts of this `relational' program \cite{L, M, BB82, FileR, ABook} work for any (carrier space) dimension $d$, point-or-particle number $N$, 
and indeed for a wide range of possible groups of structure to be stripped away, $G$ \cite{AMech, PE16}.  
This applies to the relative Lagrange coordinates and relative Jacobi coordinates sketched for $N = 3$ in Figs \ref{Jac-Med-Ineq-Fig-2}.b)-c).  
On the other hand, subsequent parts of this relational program start to depend on $d$, $N$ and $G$. 
Restricting to the current paper's use of $G = Sim(d)$ -- the similarity group -- hence the mention of carrier space origin 0, axes $A$ and scale $S$ -- 
$d$ = 1 and 2 turn out to be systematically amenable no matter what $N$ is, with $N = 3$ moreover offering further simplifications as compared to $N > 3$ in 2-$d$. 
Some of these simplifications start with Fig \ref{Jac-Med-Ineq-Fig-2}.d)'s variables.  
%
{            \begin{figure}[!ht]
\centering
\includegraphics[width=1.0\textwidth]{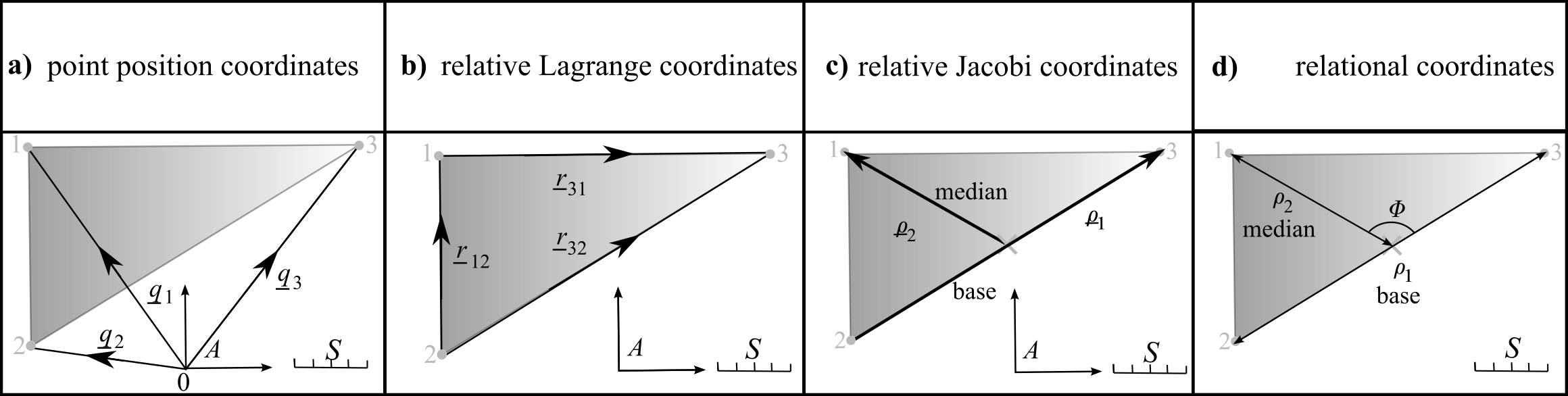}
\caption[Text der im Bilderverzeichnis auftaucht]{        \footnotesize{a) Point particle position coordinates.
b) Relative Lagrange separation vectors.   
c) Relative Jacobi separation vectors; the cross denotes the centre of mass of particles 2 and 3. 
d) Coordinates not depending on the absolute axes either: Jacobi magnitudes and the angle between them. 
To finally not depend on the scale, take the ratio $\rho_2/\rho_1$ of the Jacobi magnitudes alongside this angle. 
On the shape sphere, moreover, $\Phi$ plays the role of polar angle and the arctan of this ratio plays the role of azimuthal angle.} }
\label{Jac-Med-Ineq-Fig-2}\end{figure}            }

\mbox{ }

\ni These variables can furthermore be repackaged as a ratio of Jacobi magnitudes alongsise the relative Jacobi angle. 
The former can additionally be cast as a standard azimuthal angle (Sec 7), 
which is one of the ways of demonstrating the $Sim(2)$ shape space of triangles is revealed to be a sphere at both the topological and metric levels.
See Fig \ref{S(3, 2)-Intro} for some decor thereupon: where some of the shapes of triangles in space can be found as points within shape space.  
%
{            \begin{figure}[!ht]
\centering
\includegraphics[width=0.45\textwidth]{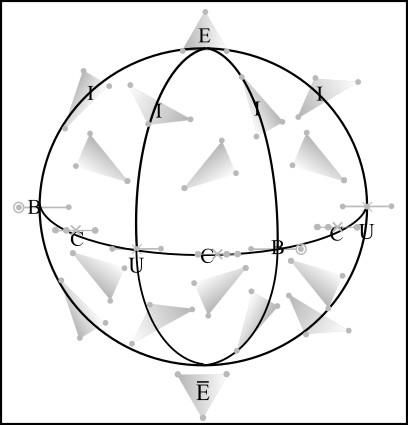}
\caption[Text der im Bilderverzeichnis auftaucht]{        \footnotesize{a) The triangleland shape sphere \cite{Kendall89, +Tri, FileR, ABook, III}  
Equilateral triangles E are at its poles, whereas collinear configurations C form its equator.
There are 3 bimeridians of isoscelesness I corresponding to 3 labelling choices for the vertices.
$C \, \cap I$ gives 3 binary collisions B and 3 uniform collinear shapes U.} }
\label{S(3, 2)-Intro} \end{figure}          }

\mbox{ }

\ni Kendall's shape sphere is an important and illustrative prototype of shape space, with the consequent Shape Theory being a highly applicable subject enjoying a current period 
of rapid theoretical and foundational growth \cite{Kendall84, Kendall89, Small, Kendall, Bhatta, JM00, HsiangStraume, FORD, 08I, 08II, GT09, +Tri, FileR, Bhatta, QuadI, AMech, MIT, 
DM16, PE16, KKH16, ABook, I, II, III, A-Pillow}.  
See e.g.\ \cite{Bookstein, Grenander96, Geom-Search-Engine, Grenander07, Younes10, Morphometrics} for further approaches.

\mbox{ }

\ni Both the 3-body problem approach in terms of relative Jacobi coordinates, 
and its Shape Theory sequel give moreover reasons to consider medians of a triangle on an equal footing with the sides. 
Together, the sides and medians constitute the Jacobi coordinate magnitudes, 
which are adapted not only to the 3-body problem but to Shape Theory as well \cite{I, III, A-Pillow, A-Pillow-2}.  
This gives a first Jacobian sense in which the medians are to be treated as coprimary to the sides.

\mbox{ }

\ni Let us also comment that the shape(-and-scale) spaces of triangles are lucidly accessible using Hopf coordinates 
(intimately related to the Hopf map and the Hopf bundle structure \cite{Hopf, Dragt, Iwai87, Nakahara, Frankel, FileR, III, A-Monopoles}).    
These provide natural Cartesian axes for the ambient $\mathbb{R}^3$ for the shape sphere $\mathbb{S}^2$. 
These axes' shape-theoretic significances are {\it 4 $\times$ mass-weighted area}, {\it anisoscelesness} and {\it ellipticity}.   
The first and last of these axes point respectively through the equilateral triangles, and through the B and U shapes described in Fig {S(3, 2)-Intro}.
The planes perpendicular to each of these in turn contain collinear shapes -- separating -- clockwise and anticlockwise labelled hemispheres of triangles,  
                                                          isosceles shapes -- separating left- and right-leaning hemispheres of scalene triangles, 
													  and regular   shapes:   separating tall and flat hemispheres of triangles.  
Note on the one hand that the equilaterality, collinearity and area complex of notions are labelling choice independent. 
On the other hand, as formulated in \cite{+Tri, FileR, III} isoscelesness, left- and right-leaning, regularness, tallness and flatness depend on such a choice.
This Hopfian description is moreover sides-to-medians symmetric \cite{2-Herons}, giving a second motivation to treating sides and medians on the same footing.  
I subsequently found \cite{2-Herons} that the Jacobi and Hopf formulations provide successively clear renditions of Heron's formula:  
mass weighted sides-to-medians symmetric versions and a diagonal formulation that gives the Hopf coordinates {\sl and} forms the on-sphere condition 
that re-derives Kendall's Theorem that the space of triangles is a sphere.  
Ellipticity and anisoscelesness were furthermore shown \cite{2-Herons} to be eigenvectors shared by the Heron map and the sides-to-medians involution.
%
{            \begin{figure}[!ht]
\centering
\includegraphics[width=1.0\textwidth]{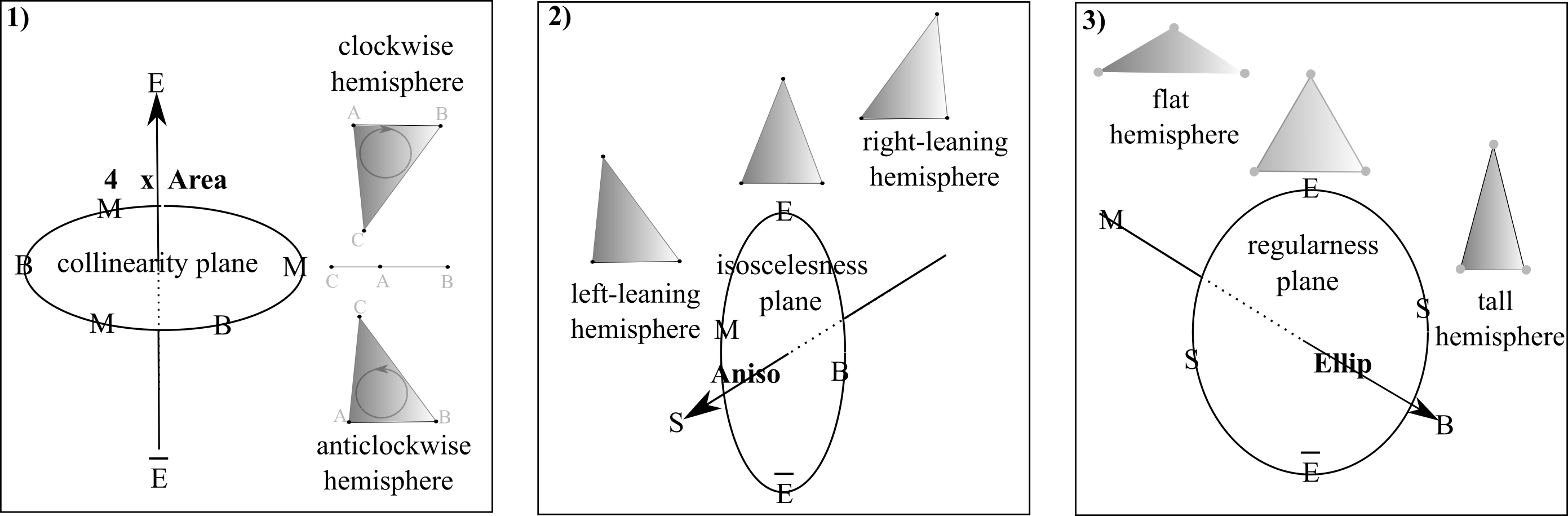}
\caption[Text der im Bilderverzeichnis auftaucht]{        \footnotesize{1) Collinear shapes separate hemispheres of clockwise and anticlockwise triangles. 

\mbox{ }

\ni 2) Isosceles triangles separate hemispheres of left-leaning and right-leaning triangles.

\mbox{ }

\ni 3) Regular triangles separate hemispheres of tall and flat triangles.} }
\label{Hopf-Axes} \end{figure}          }

\mbox{ }

\ni Moreover, considering right rather than regular decor, Lewis Carroll's pillow problem \cite{Pillow} of what is Prob(obtuse) 
has been given a solid and simple shape-theoretic answer \cite{MIT, III, A-Pillow} with numerous further similar problems rendered accessible \cite{A-Pillow, A-Pillow-2, IV}.   

\mbox{ }

\ni Returning to consideration of talless and flatness, Kendall termed very tall or very flat triangles {\it splinters}; 
see Fig \ref{S(3, 2)-Splinters}.a) for the diversity of types of splinter, and Fig \ref{S(3, 2)-Splinters}.b) for where these are located in the shape sphere. 
Kendall furthermore devised \cite{Kendall89, Kendall} statistical tests for splinters as configurations of approximately-collinear triples of points, 
by which splinters are a topic of considerable shape-theoretic significance.  
%
{            \begin{figure}[!ht]
\centering
\includegraphics[width=0.66\textwidth]{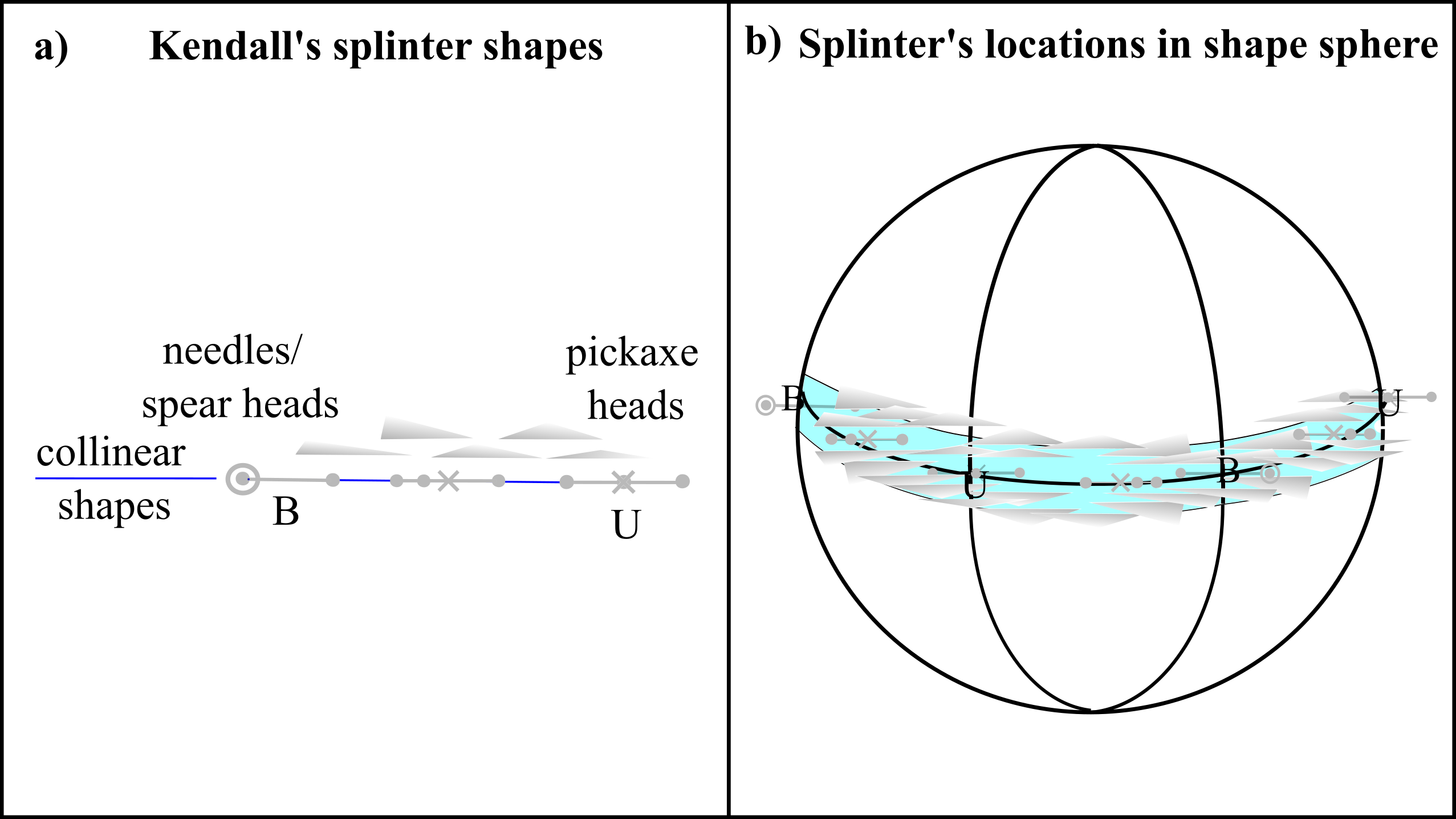}
\caption[Text der im Bilderverzeichnis auftaucht]{        \footnotesize{a)  Types of splinter shapes, and b) where they are to be found in the triangleland shape sphere.
%
} }
\label{S(3, 2)-Splinters} \end{figure}          }

\mbox{ }

\ni{\bf Outline of the rest of this paper}

\mbox{ }

\ni Sec 2 provides preliminaries about sides, medians and radii of circles associated with the triangle.
We also review some well-known inequalities bounding the lengths of medians \cite{Court, IMO-0, IMO, IMO-2}; perimeter, inradius and circumradius enter these bounds.  
We next provide ratio versions of these quantities and inequalities in Sec 3. 
In Sec 4 and 5, we proceed to outline relative Jacobi vectors and separations, including mass-weighted versions of these.  
This permits us to reformulate the preceding inequalities as bounds on mass-weighted relative Jacobi magnitudes in Sec 6: {\it Jacobi inequalities}.   
This work identifies the $4/3$ -- powers of which occur frequently in the theory of medians -- as the ratio of Jacobi masses.

\mbox{ }

\ni We outline the shape sphere and the relational space of scaled triangles in Sec 7, and the Hopf coordinates presentation of these in Sec 8.
We consider regular, tall and flat triangles in further detail in Sec 9, and the notion of democratic, i.e.\ clustering independent alias labelling independent notions in Sec 10.
Democratized ellipticity moreover trivializes, due to ellipticity being the difference of base-side and median second moments, while the averages of these are equal to each other.  
Because of this, in Sec 11 we introduce a distinct `linear ellipticity' quantifier of tallness-or-flatness of triangles whose democratization is nontrivial, 
and find inequalities on this. 
Some of this paper's inequalities are shape-independent bounds, whereas others' bounds depend on the isoperimetric ratio and arithmetic-to-geometric side mean ratio shape variables. 

\vspace{10in}

\section{Preliminary definitions and inequalities}
%
{            \begin{figure}[!ht]
\centering
\includegraphics[width=0.57\textwidth]{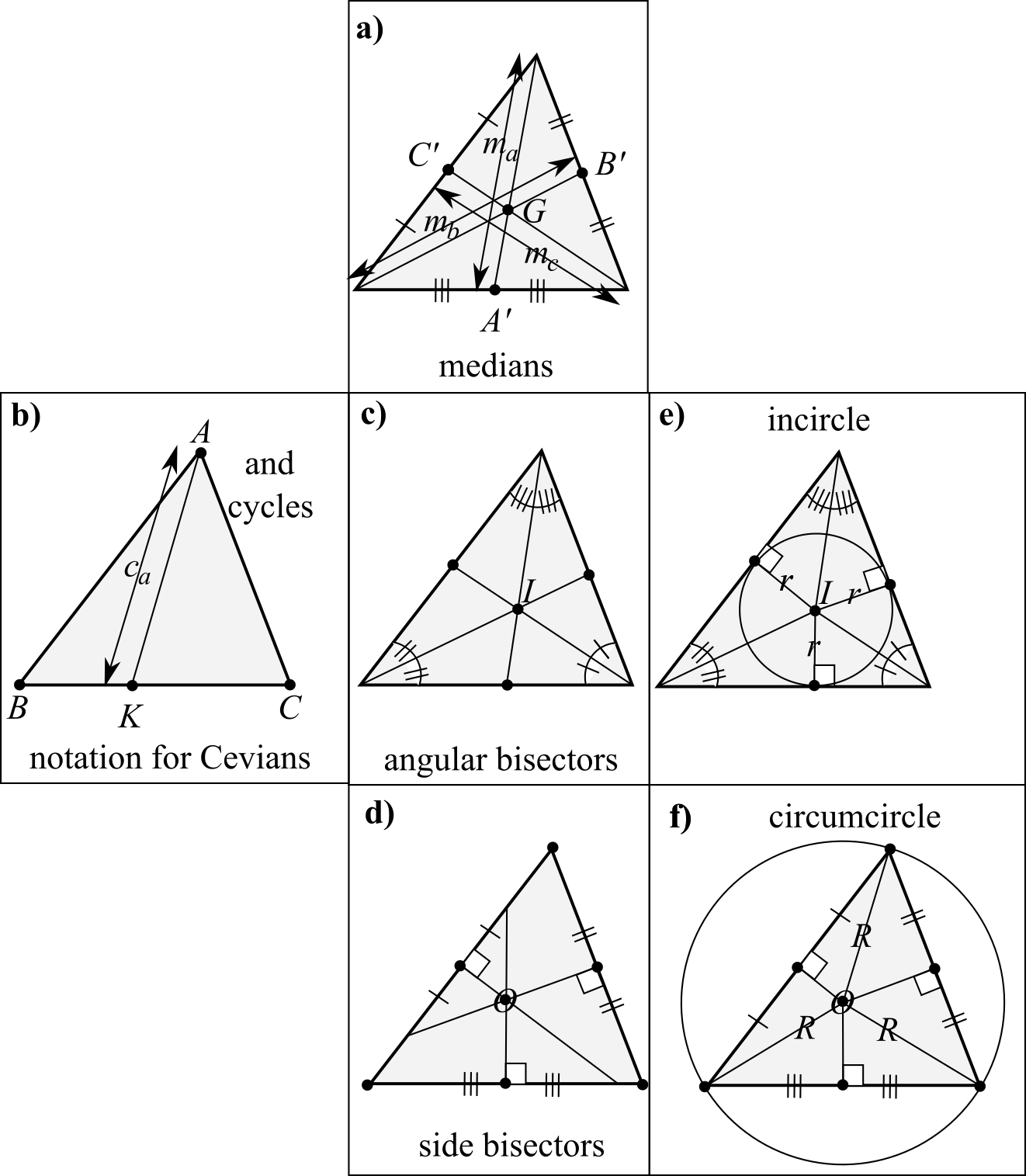}
\caption[Text der im Bilderverzeichnis auftaucht]{        \footnotesize{a) Definition of the medians; these concur at the {\it centroid}, alias {\it centre of mass}, $G$.
b) General Cevian set-up, of which medians and internal angle bisectors are a particular subcase. 
c) The internal angle bisectors $h_a$ from each vertex to its opposite side. 
d) The sides' perpendicular bisectors; note that these in general do not go through the vertex opposite the side. 
e) The angle bisectors moreover define the incircle as indicated, with incentre $I$ and inradius $r$. 
f) The perpendicular bisectors of the sides moreover define the circumcircle as indicated, with circumcentre $O$ and circumradius $R$. }  } 
\label{Med-In-Circum} \end{figure}           }

\ni{\bf Definition 1} Consider an arbitrary triangle $\triangle\,ABC$, denoted as in \ref{Triangle-Notation}, with {\it medians} as per \ref{Med-In-Circum}.a).  
It will also be useful for us to use $m_i$ to denote $m_a$, $m_b$, $m_c$.  

\mbox{ } 

\ni{\bf Remark 1} As we are treating the sides and the medians on an equal footing, we have more definitions (or at least accordances of equal significance) 
than in hitherto standard treatments of triangles. 

\mbox{ } 

\ni{\bf Definition 2} Three such pairs of quantities used in this paper are defined in Fig \ref{6-variables}, including various rescalings.

\mbox{ } 

\ni{\bf Remark 2} We show later that the middle pair are moreover in fact proportional to each other, so these are in fact five independent quantities.  
%
{            \begin{figure}[!ht]
\centering
\includegraphics[width=0.85\textwidth]{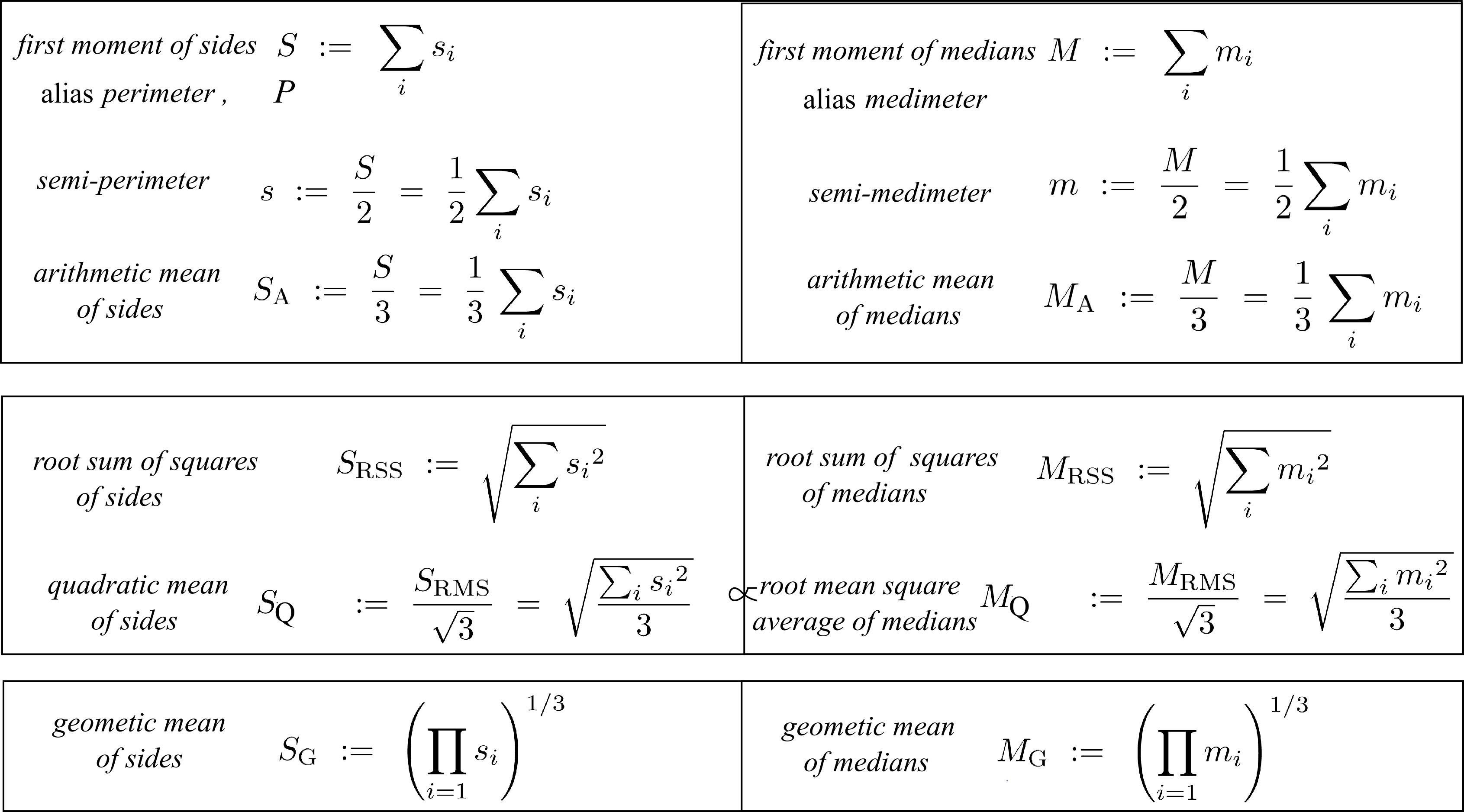}
\caption[Text der im Bilderverzeichnis auftaucht]{        \footnotesize{Three pairs of variables, with one inter-relation and various useful and conceptually meaningful rescalings. 
Note that we are using the letters 'M' and 'S' to denote quantites which are dimensionally lengths, while also keeping track of which are side-based and which are median-based.}  } 
\label{6-variables} \end{figure}           }

\mbox{ }

\ni{\bf Definition 3} We also use $\mbox{Area}(ABC)$, or $\mbox{Area}$ for short when unambiguous, to denote the area of $\triangle\,ABC$.
It turns out that this also admits a useful and geometrically significant rescaling, namely the {\it Tetra-Area} 
\be 
T := 4 \times Area \mbox{ } . 
\label{Tetra-Area}
\ee 
\ni{\bf Remark 3} Computationally, $Area$ is given by 
\be
Area = a \, b \, \mbox{sin} \, \gamma \mbox{ } \mbox{ or cycles }
\ee 
in terms of s-a-s (side-included-angle-side) data, or {\it Heron's formula}
\be
Area = \sqrt{s(s - a)(s - b)(s - c)} \mbox{ }  
\label{Heron}
\ee
in terms of s-s-s (3-sides) data.  
The latter also admits the expanded form 
\be
T^2 \mbox{ } = \mbox{ } (2 \, a^2 b^2 - c^4) + \mbox{ cycles } \mbox{ } , 
\label{Expanded-Heron}
\ee 
which we shall subsequently further reformulate.  

\mbox{ }

\ni{\bf Definition 4} We furthermore use $l_i$, $i = 1$ to $3$ to denote {\it internal angle bisectors} (Fig \ref{Med-In-Circum}.d).  

\mbox{ }

\ni{\bf Remark 4} In the current paper, we make much less use of Definition 4's concepts than Definition 1 to 3's.
This lesser use is moreover standard, through these entering the definitions of the following radii of circles that can be associated with our original triangle. 

\mbox{ }

\ni{\bf Definition 5} The {\it inradius} $r$, and {\it circumradius} $R$ are defined as per \ref{Med-In-Circum}.e) and f) respectively.  

\mbox{ }

\ni{\bf Lemma 1} i) 
\be
Area = r \, s \mbox{ } .  
\label{Area-r-s}
\ee
ii) 
\be
Area = r_a (s  - a) \mbox{ } \mbox{ and cycles } .    
\label{Area-ra-s}
\ee
iii) 
\be
S_{\sG}\mbox{}^3 = abc = 4 \, s \, r \, R  = 4 \, Area \, R  = R \, T  \mbox{ } .  
\ee 
\ni{\underline{Proof}} These are all standard; see e.g.\ \cite{Coxeter} for i), ii), as well as obtaining Heron's formula (\ref{Heron}) from ii).  
The first equality of iii) is Fig \ref{6-variables}.e)'s definition, 
the second derived in e.g.\ \cite{Johnson}, 
the third follows from i) and 
the new fourth equality form from (\ref{Tetra-Area}). $\Box$ 

\mbox{ }

\ni{\bf Remark 5} Various either simple or well-known inequalities for side and median length quantities are as follows.

\mbox{ }

\ni{\bf Lemma 2 (Preliminary crude joint bounds on sides and medians)} i) 
\be
2 \,  r    \mbox{ } <  \mbox{ }       s_i                 \mbox{ } , \mbox{ } \mbox{ }      m_i                      \mbox{ } \leq 2 \, R \mbox{ } . 
\ee 
ii) 
\be
6 \,  r    \mbox{ } <  \mbox{ }        P                  \mbox{ } , \mbox{ } \mbox{ }      M                        \mbox{ } \leq 6 \, R \mbox{ } .
\ee 
iii)
\be 
12 \ r^2 \mbox{ } \leq  \mbox{ } S_{\sR\sS\sS}\mbox{}^2   \mbox{ } , \mbox{ } \mbox{ }      M_{\sR\sS\sS}\mbox{}^2   \mbox{ } \leq 12 \, R \mbox{ } .
\ee 
\ni{\underline{Proof}} i) For the first inequality, sides and medians both exceed the diameter $2 \, r$ of the inscribed circle.
For medians, this is from comparing the diameter perpendicular to the side with the portion of the in-general slanting median up to the same height away from the side.
Even when the median runs along this -- isosceles triangles -- the inequality is strict.
For sides, this is from projecting the circle onto each side, giving a diameter-wide shadow which still fails to cover the two ends of the side.  

\mbox{ }

\ni For the second inequality, sides and medians are both segments inside the circle, so their maximal size is $2 \, R$.
For sides, this is attained with equality iff the triangle is right, the right side in question then being the hypotenuse.
For medians, however, this inequality is strict.  

\mbox{ }

\ni  ii) and iii) then follow from summing i), and the squares of i), respectively, over all 3 sides or all 3 medians.  $\Box$ 

\mbox{ }

\ni{\bf Remark 6} ii) and iii) are mostly provided for comparison with subsequent sharper inequalities, which do discern between sides and medians.   
Many of these can be accessed via the median being a subcase of the following concept, for which general theorems follow.  

\mbox{ }

\ni{\bf Definition 6} The segment $AK$ from a triangle's vertex $A$ to a point $K$ on the opposite side $BC$ is known as a {\it Cevian}.  

\mbox{ }

\ni{\bf Remark 7} See Fig \ref{Med-In-Circum}.b) for Cevians $c_a$;  medians $m_a$ are clearly a subcase.    

\mbox{ }

\ni{\bf Remark 8} The ratios 
\be 
\frac{KC}{BC} \mbox{ } \mbox{ and } \mbox{ } \frac{BK}{BC} 
\ee
have affine significance, entering Ceva's own theorem \cite{PS70, Silvester} about concurrence of Cevians; 
medians themselves obviously concur e.g.\ by simpler centre-of-mass arguments.  
In Euclidean geometry, moreover, Cevians have lengths as well as affine properties, and it is in fact the following `Cevian length theorem' that we make use of.  

\mbox{ }

\ni{\bf Lemma 3 (Stewart's Theorem)} Let $\triangle ABC$ be a triangle with $K$ an arbitrary point on side $BC$.  
Then
\be
{AK}^2 \mbox{ } = \mbox{ } \frac{KC}{BC} \, {AB}^2 \mbox{ } + \mbox{ } \frac{BK}{BC} \, {AC}^2 - BK \, KC \mbox{ } .  
\ee 
\ni{\underline{Proof}} See e.g.\ \cite{PS70}. $\Box$ 

\mbox{ }

\ni{\bf Corollary 1} i) The median lengths' squares are given by 
\be
m_a\mbox{}^2 \mbox{ } = \mbox{ } \frac{2 \, b^2 + 2 \, c^2 - a^2}{4} \mbox{ } \mbox{ and cycles } .
\label{0.1}
\ee
\ni ii) The quadratic means of sides and of medians are related by 
\be 
S_{\sQ} \mbox{ } = \mbox{ } \frac{2}{\sqrt{3}} \, M_{\sQ} 
                 =                \kappa       \, M_{\sQ} \mbox{ } , 
\label{0.2}
\ee
for 
\be 
\kappa := \frac{2}{\sqrt{3}} \mbox{ } . 
\ee 
\ni{\underline{Proof}} i)  This readily follows from Stewart's Theorem, as per Problem 1 of \cite{IMO}. 

\mbox{ }

\ni ii) then follows immediately from summing i) over all cycles, and square-rooting. $\Box$

\mbox{ }

\ni{\bf Remark 9} $\kappa$ is for now to be treated as a number, powers of which are recurrent in the geometrical theory of the triangle. 
A conceptual and physical meaning for $\kappa$ will moreover be elucidated in Sec 4. 

\mbox{ }

\ni{\bf Structure 1} i) In Linear Algebra form \cite{2-Herons}, 
\be
             \left( \stackrel{  \stackrel{  \mbox{$m_a\mbox{}^2$}  }{  \mbox{$m_b\mbox{}^2$}  }  }{\mbox{$m_c\mbox{}^2$}} \right) \mbox{ } = \mbox{ } 
\frac{1}{4}  \left( \stackrel{  \stackrel{  \mbox{$   -              1  \mbox{ } \mbox{ } \mbox{ }   2 \mbox{ } \mbox{ }  \mbox{ } \mbox{ } \,  2$}  }  
                                         {  \mbox{$\mbox{ } \mbox{ } \,   2  \mbox{ }    -               1 \mbox{ } \mbox{ } \mbox{ }  2$}  }  }
										 {  \mbox{$\mbox{ } \mbox{ } \, 2  \mbox{ } \mbox{ } \mbox{ }  \mbox{ } 2 \mbox{ }    -               1$}  } \right) 
             \left( \stackrel{\stackrel{\mbox{$a^2$}}{\mbox{$b^2$}}}{\mbox{$c^2$}} \right) \mbox{ } , 
\label{0.3}
\ee
i.e.\ 
\be
m_i\mbox{}^2 \mbox{ } = \mbox{ } \frac{1}{4} \, A_{ij} s_j^2 
\label{0.4}
\ee
for 
\be
\underline{\underline{A}} \mbox{ } := \mbox{ } 
\left( \stackrel{  \stackrel{  \mbox{$   -              1  \mbox{ } \mbox{ } \mbox{ }  2 \mbox{ } \mbox{ } \mbox{ } \mbox{ } \,  2$}  }  
                            {  \mbox{$\mbox{ } \mbox{ } \mbox{ } 2  \mbox{ }    -               1 \mbox{ } \mbox{ }  \mbox{ } \mbox{ } \,  2$}  }  }
							{  \mbox{$\mbox{ } \mbox{ } \mbox{ } 2  \mbox{ } \mbox{ } \mbox{ }  \mbox{ } 2 \mbox{ }    -               1$}  } \right)
\label{0.5}
\ee
\ni ii) Inverting, 
\be
s_i\mbox{}^2 \mbox{ } = \mbox{ } \frac{4}{9} \, A_{ij} m_j^2 \mbox{ } . 
\label{0.6}
\ee
\ni{\bf Remark 10} That the same $A_{ij}$ appears in the inverted expression indicates that $A_{ij}$ is proportional to an {\it involution} $J_{ij}$, i.e.\ it is a matrix such that 
\be
\underline{\underline{J}}^2 = \underline{\underline{1}}: \mbox{ the identity matrix } .  
\label{0.7}
\ee
We can thus further tidy up Corollary 1's Linear Algebra formulation of the sides---medians relation by identifying and using $\underline{\underline{J}}$, 
as follows \cite{2-Herons}. 

\mbox{ }

\ni{\bf Corollary 2} \cite{2-Herons} i) 
\be
m_i\mbox{}^2 = \kappa^{-2} \, J_{ij} s_j^2 \mbox{ } , 
\label{0.8}
\ee
for involution 
\be
\underline{\underline{J}} \mbox{ } := \mbox{ } 
\frac{1}{3}   
\left( \stackrel{  \stackrel{  \mbox{$   -              1  \mbox{ } \mbox{ } \mbox{ }  2 \mbox{ } \mbox{ } \mbox{ } \,  2$}  }  
                                         {  \mbox{$\mbox{ } \mbox{ } \, \, 2  \mbox{ }    -               1 \mbox{ } \mbox{ }  \mbox{ } \,  2$}  }  }
										 {  \mbox{$\mbox{ } \mbox{ } \, 2  \mbox{ } \mbox{ } \mbox{ }  2 \mbox{ }    -               1$}  } \right)             \mbox{ } .
\ee
ii) Inverting, 
\be
s_i\mbox{}^2 = \kappa^2 \, J_{ij} m_j^2 \mbox{ } .   
\label{0.9}
\ee
\ni{\bf Lemma 4 (Perimeter bounds on the sum of the medians $M$} \cite{Court}.  
\be
\kappa^{-2} S \mbox{ } \leq \mbox{ } M 
              \mbox{ } \leq \mbox{ } S \mbox{ } . 
			  \label{Ineq-1}
\ee 
\ni{\underline{Proof}} This results from combining worked problem 2's of \cite{IMO-0} sharper lower bound with worked problem 1's upper bound.  $\Box$

\mbox{ }

\ni{\bf Remark 11} Including degenerate triangles -- as is required in Shape Theory -- means that indeed the saturated version of these inequalities is required, 
with the value 1 occurring at the binary coincidence-or-collision $B$ and $\kappa^{-2}$ occurring for the uniform collinear configuration $U$. 

\mbox{ }

\ni{\bf Lemma 5} The root sum of squares is bounded by the square of the semi-perimeter length, $s$, according to 
\be
s^2 \mbox{ } \leq \mbox{ } \kappa^{-2} \, S_{\sR\sM\sS}\mbox{}^2  
    \mbox{ }   =  \mbox{ }                M_{\sR\sM\sS}\mbox{}^2                                 \mbox{ } .
\label{0.10}
\ee 
\ni{\underline{Proof}} (along the lines of worked Problem 5 of \cite{IMO})
\be
s^2 \mbox{ } = \mbox{ } \frac{1}{4} S^2 
    \mbox{ } = \mbox{ } \frac{1}{4} \left(  S_{\sR\sM\sS}  + 2 \sum \sum_{i>j} s_i s_j  \right)  \mbox{ } . 
\label{0.11}
\ee
Also  
\be
0 \mbox{ } \leq \mbox{ } (a - b)^2 
  \mbox{ }   =  \mbox{ }  a^2 + b^2 - 2 \, a \, b \mbox{ } \mbox{ and cycles } .  
\label{0.12}
\ee
Thus
\be
s^2 \mbox{ } \mbox{ } \leq \mbox{ } \mbox{ } \frac{1}{4}\left(  S_{\sR\sM\sS}\mbox{}^2  + 2 \, S_{\sR\sM\sS}\mbox{}^2  \right) 
             \mbox{ }   =  \mbox{ }          \kappa^{-2}     \, S_{\sR\sM\sS}\mbox{}^2 
			 \mbox{ }   =  \mbox{ }                             M_{\sR\sM\sS}\mbox{}^2                                  \mbox{ } , 
\label{0.13}
\ee
where we have used Corollary 1.ii) in the last step.  $\Box$

\mbox{ }

\ni{\bf Structure 2} Let us now reformulate the tetra-area squared as the Heron form 
\be 
T^2     =  H_{ij} s_i\mbox{}^2 s_i\mbox{}^2 \mbox{ } . 
\ee 
This is a quadratic form whos inputs are moreover squares. 
The matrix involved is the {\it Heron matrix} 
\be
\underline{\underline{H}} \mbox{ } := \mbox{ } 
\left( \stackrel{  \stackrel{  \mbox{$   -              1  \mbox{ } \mbox{ } \mbox{ }  1 \mbox{ } \mbox{ } \mbox{ } \mbox{ } \,  1$}  }  
                                         {  \mbox{$\mbox{ } \mbox{ } \mbox{ } 1  \mbox{ }    -               1 \mbox{ } \mbox{ }  \mbox{ } \mbox{ } \,  1$}  }  }
										 {  \mbox{$\mbox{ } \mbox{ } \mbox{ } 1  \mbox{ } \mbox{ } \mbox{ }  \mbox{ } 1 \mbox{ }    -               1$}  } \right)
\label{0.5-b}
\ee
In \cite{2-Herons}, I showed that furthermore $\underline{\underline{H}}$ and $\underline{\underline{J}}$ commute, 
underlying the similarity between the sides and medians versions of Heron's formula.
The latter takes the Linear Algebra form 
\be 
T^2 = \kappa^2 \, H_{ij} m_i\mbox{}^2 m_i\mbox{}^2 \mbox{ } , 
\ee 
thus only being out by a factor of $\kappa^2$ from the sides form.  

\mbox{ }

\ni{\bf Lemma 6 (Inradius--circumradius bounds)} i) The medimeter is bounded according to  
\be
r \mbox{ } \leq \mbox{ } \frac{M}{9} 
  \mbox{ } \leq \mbox{ } \frac{R}{2}                                                        \mbox{ } . 
\label{0.14}
\ee
ii) The sum squared of medians is bounded according to 
\be 
r^2 \mbox{ } \leq \mbox{ } \frac{  M_{\sR\sM\sS}\mbox{}^2  }{  27  } 
    \mbox{ } \leq \mbox{ } \frac{            R^2           }{   4  }                        \mbox{ } .  
\label{0.15} 
\ee
\ni{\underline{Proof}} This is a subset of worked Problems 1 and 2 of \cite{IMO-2}. $\Box$

\mbox{ }

\ni{\bf Remark 12} On the one hand,    i) tightens Lemma 2.ii) from $6 \, r$ to $9 \, r$ -- a factor of 3/2 -- and from $6 \, R$ to $9 \, R/2$: a factor of 4/3. 

\mbox{ }

\ni               On the other hand, ii) tightens Lemma 2.iii) from $12 \, r$ to $27 \, r$ --  a factor of $9/4  = (3/2)^2$ -- 
                                                           and from $12 \, R^2$ to $27 R^2/4$: a factor of $16/9 = (4/3)^2$.

\mbox{ }

\ni{\bf Remark 13} Inradius $r$ and circumradius $R$ are moreover not natural or primary constructs from a shape-theoretic point of view, 
being `extrinsic' features of triangles, but we can, rather, reformulate these quantities as per Lemma 1.i) and iii), giving the following further form for the inequalities.

\mbox{ }

\ni{\bf Corollary 3} i) 
\be
\frac{T}{S}                \mbox{ } <  \mbox{ }          s_i     \mbox{ } , \mbox{ } \mbox{ }  
                                                         m_i     \mbox{ } , \mbox{ } \mbox{ }  
		    											 S_A     \mbox{ } , \mbox{ } \mbox{ }  
													     M_A     \mbox{ } , \mbox{ } \mbox{ } \leq \mbox{ } 2 \, \frac{S_{\sG}\mbox{}^3}{T}                 \mbox{ } . 
\ee  
ii)
\be 
\left(\frac{T}{S}\right)^2  \mbox{ } < \mbox{ }       S_{\sQ}\mbox{}^2 ,  M_{\sQ}\mbox{}^2 
                            \mbox{ } \leq \mbox{ }  4 \, \left(\frac{S_{\sG}\mbox{}^3}{T}\right)^2                                                          \mbox{ } .
\ee
\ni{\bf Corollary 4}
\be
\sqrt{3} \, \frac{T}{S} \mbox{ } \leq \mbox{ } \kappa \, M_{\sA}    \mbox{ } \leq \mbox{ }   \sqrt{3} \, \frac{S_{\sG}\mbox{}^3}{T}                                                           \mbox{ } . 
\label{0.14-b}
\ee
ii) The second moment of medians is bounded according to 
\be 
3 \left(\frac{T}{S}\right)^2   \mbox{ } \leq \mbox{ }              S_{\sQ}\mbox{}^2 
                                          =            \kappa^2 \, M_{\sQ}\mbox{}^2 
                               \mbox{ } \leq \mbox{ }  3 \left(\frac{  S_{\sG}\mbox{}^2  }{  T  }\right)^2                                                  \mbox{ } .  
\label{0.15-b} 
\ee
\ni{\bf Remark 14}  These two inequalities furthermore admit the following joint repackaging.

\mbox{ }

\ni{\bf Theorem 1}
\be
\sqrt{3} \, \frac{T}{S} \mbox{ } \leq \mbox{ } \kappa \, M_{\sA} 
                     \mbox{ } \leq \mbox{ }    \kappa \, M_{\sQ} 
					            =                        S_{\sQ}
                     \mbox{ } \leq \mbox{ } \sqrt{3} \, \frac{S_{\sG}\mbox{}^3}{T}                                                                            \mbox{ } . 
\label{0.14-c}
\ee
\ni{\underline{Proof}} The middle inequality is by 
\be
M^2 - 3 \, M_{\sR\sS\sS}\mbox{}^2  = M_{ij}m_im_j \mbox{ } \leq \mbox{ }  0 \mbox{ } ,
\ee
for 
\be
\underline{\underline{M}} \mbox{ } := \mbox{ } 
\left( \stackrel{  \stackrel{  \mbox{$ - 2  \mbox{ } \mbox{ } \mbox{ }  1 \mbox{ } \mbox{ } \mbox{ } \, 1$}  }  
                                         {  \mbox{$\mbox{ } \mbox{ } \, \, 1  \mbox{ } -2 \mbox{ } \mbox{ }  \mbox{ } \,  1$}  }  }
										 {  \mbox{$\mbox{ } \mbox{ } \, 1  \mbox{ } \mbox{ } \mbox{ }  1 \mbox{ }    - 2$}  } \right)             \mbox{ } .
\ee
and where the inequality follows from the negative-definiteness of this matrix.
The rest follows from Corollary 4. $\Box$

\section{Ratio reformulation of triangles and inequalities}
%
{            \begin{figure}[!ht]
\centering
\includegraphics[width=0.65\textwidth]{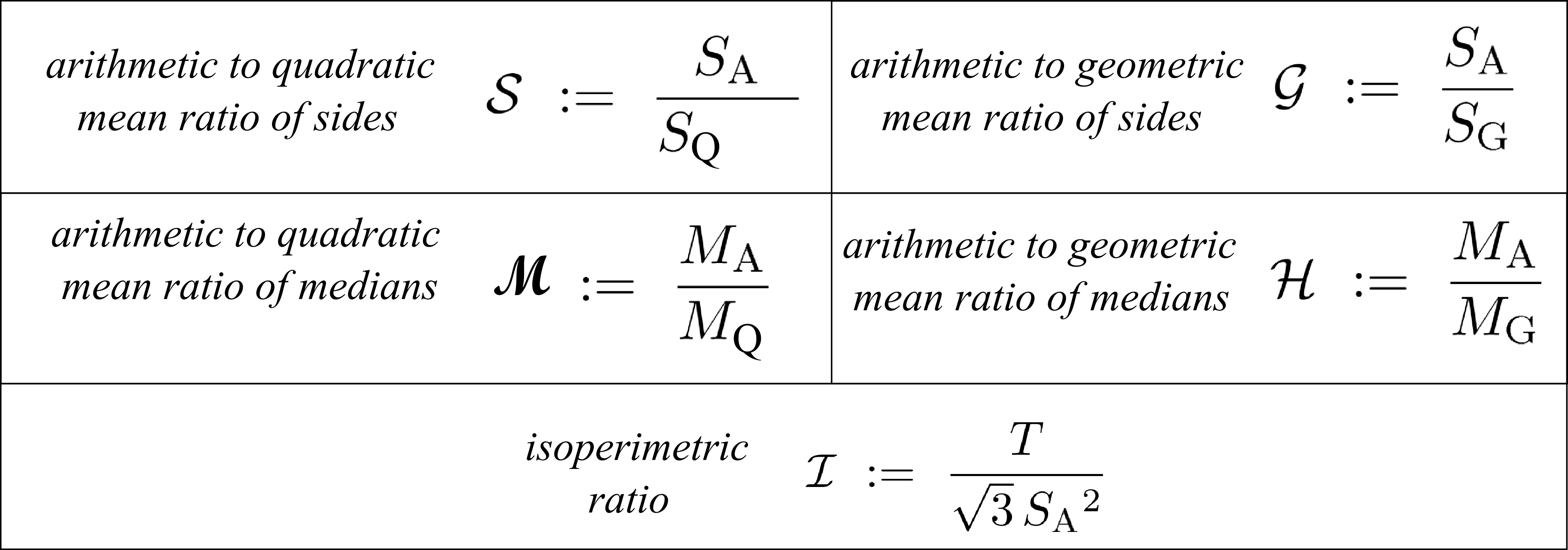}
\caption[Text der im Bilderverzeichnis auftaucht]{        \footnotesize{A choice of five independent ratios, as supported by $Area$ and Figure \ref{6-variables} 
and five other independent quantities.}   } 
\label{5-ratios} \end{figure}           }

We started with seven quantities, but showed two to be equal up to proportion.
Thus five independent ratios are supported.
For reasons which later become clear, we pick two side-quantity ratios, two median-quantity ratios and the isoperimetric ratio for this purpose, as follows.  

\mbox{ }

\ni{\bf Definition 7} The five independent ratios the current paper conceives in terms of are given in Fig \ref{5-ratios}.  
  
\mbox{ }

\ni{\bf Remark 15} These are all scaled to have range 0 to 1.  
Setting this up requires finding and evaluating their maxima (see \cite{A-Perimeter} for this and further extremization calculations.  

\mbox{ }

\ni{\bf Remark 16} To carry over to Shape Theory, we first recast all the inequalities in Sec 2 in terms of ratios.  
We first define the quantities to appear in these ratios, which are composites of some of the above five `most natural' ratios.  

\mbox{ }

\ni{\bf Definition 8} 
\be 
{\cal F} \mbox{ } := \mbox{ } \frac{M}{S} \mbox{ } = \mbox{ }  \frac{M_{\sA}}{S_{\sA}} 
\ee 
is the {\it medimeter to perimeter ratio}.  
%

\mbox{ }

\ni{\bf Remark 17} 
\be 
\mbox{ } \frac{M_{\sQ}}{S_{\sQ}}
\ee 
is sterile by the constancy implied by (\ref{7}).  
The following alternative quantity, however, does have nontrivial shape-theoretic content (and is what occurs in the below inequalities). 

\mbox{ }

\ni{\bf Definition 9} The second ratio featuring in our inequalities is 
\be
{\cal L} \mbox{ } := \mbox{ } \frac{  M_{\sQ}  }{  S_A  } : 
\ee 
the {\it quadratic-to-arithmetic mean ratio of medians}.

\mbox{ }

\ni{\bf Definition 10} The third and fourth ratios featuring in the ratio version of the traditional form of the inequalities are 
\be 
\scN     \mbox{ } := \mbox{ } \frac{  r  }{  S  } \mbox{ } : 
\label{n}
\ee
the {\it inradius per unit perimeter}, and  
\be 
{\cal N} \mbox{ } := \mbox{ } \frac{  R  }{  S  }  \mbox{ } : 
\label{N}
\ee
is the {\it circumradius per unit perimeter}.  

\mbox{ }

\ni{\bf Remark 18} Moreover, by Lemma 1.i), we evaluate (\ref{n}) to be 
\be 
\scN \mbox{ } = \mbox{ } \frac{T}{2 \, S^2} \mbox{ } , 
\ee
by which this is just proportional to the isoperimetric ratio ${\cal I}$ of Fig \ref{5-ratios}.e):  
\be 
\scN \mbox{ } = \mbox{ } \frac{{\cal I}}{  6 \sqrt{3}  } \mbox{ } . 
\ee 
Additionally, by Lemma 1.iii), we evaluate (\ref{N}) to be   
\be 
{\cal N} = \frac{S_{\sG}\mbox{}^3}{T \, S} \mbox{ } ,  
\ee 
so this is proportional to a product of our more basic ratios, 
\be
{\cal N} = \frac{1}{3\sqrt3 \, {\cal I} \, {\cal G}^3} \mbox{ } .
\ee 

\mbox{ }

\ni{\bf Lemma 7 (Rational form of the inequalities.)}
i) 
\be 
\kappa^{-2} \mbox{ } \leq \mbox{ } {\cal F}  
            \mbox{ } \leq \mbox{ } 1 \mbox{ } .
\label{Rat-Ineq-1}
\ee
ii) 
\be
\scN \mbox{ } \leq \mbox{ } \frac{  {\cal F}  }{  9  } 
     \mbox{ } \leq \mbox{ } \frac{  {\cal N}  }{  2  } \mbox{ } . 
\ee 
iii)  
\be
\scN^2 \mbox{ } \leq \mbox{ } \frac{  {\cal E}^2  }{  27  }  
       \mbox{ } \leq \mbox{ } \frac{  {\cal N}^2  }{  4  }                          \mbox{ } . 
\ee
\ni{\underline{Proof}} i) Divide Lemma 6.ii) by $P$ and use the definition of $F$.

\mbox{ }

\ni ii) Divide Lemma 6.i) by $9 \, P$ and use the definitions of ${\cal F}$, $\scN$ and ${\cal N}$.

\mbox{ }

\ni iii) Divide Lemma 6.ii) by $27 \, P^2$ and use the definitions of ${\cal F}$, $\scN$ and ${\cal N}$. $\Box$ 

\mbox{ }

\ni{\bf Remark 19} ${\cal F}$'s maximum is thus also 1 in accord with our standardization convention, but its range is $\left[\frac{3}{4}, \, 1\right]$ rather than [0, 1].
As \cite{A-Perimeter} shall show, the range for ${\cal L}$ is $\left[\frac{1}{2\sqrt{2}}, 1\right]$.
${\cal F}_{\sm\sa\sx}$ is at the binary coincidence-or-collision $B$, whereas ${\cal L}_{\sm\sa\sx}$ is at the equilateral triangle $E$.
Both ${\cal F}_{\sm\si\sn}$ and ${\cal L}_{\sm\si\sn}$ are at the uniform collinear configuration $U$. 

\mbox{ }

\ni{\bf Corollary 5} In terms of the isoperimetric and arithmetic-to-geometric mean side ratios, i)
\be
{\cal I} \mbox{ } \leq \mbox{ }        \kappa \, {\cal F}  
         \mbox{ } \leq \mbox{ } {\cal I}^{-1}    {\cal G}^{-3} \mbox{ } . 
\ee 
ii)  
\be
{\cal I}^2  \mbox{ } \leq \mbox{ }   \kappa^2  {\cal L}^2 
            \mbox{ } \leq \mbox{ }       {\cal I}^{-2} {\cal G}^{-6}  \mbox{ } . 
\ee
\ni{\underline{Proof}} Use Definition 10 and Remark 18 in Lemma 6, multiplying i) by 2 and ii) by 4. $\Box$

\mbox{ }

\ni{\bf Remark 20}  These two inequalities then furthermore admit the following joint repackaging, by use of Theorem 1.

\mbox{ }

\ni{\bf Theorem 2}  
\be
{\cal I} \mbox{ } \leq \mbox{ } \kappa   \, {\cal F}  
         \mbox{ } \leq \mbox{ } \kappa   \, {\cal L}
                    =                       {\cal S}^{-1}		 
		 \mbox{ } \leq \mbox{ }             {\cal I}^{-1} {\cal G}^{-3} \mbox{ } . 
\label{Rat-Ineq-2}
\ee 
\ni{\bf Remark 21} This is an intrinsic reconceptualization of in-and-circumradius bounds on median moments, 
to now be in terms of the isoperimetric ratio and arithmetic-to-geometric sides ratio shape quantities.

\section{Jacobi coordinates for the triangle}

\ni{\bf Structure 3} Let us now consider our triangle's vertices to be equal-mass particles with position vectors $\uq_I$ ($I = 1$ to $3$) 
relative to an absolute origin 0, axes $A$ and scale $S$ Fig \ref{Jac-Med-Ineq-Fig-2}.a).

\mbox{ }

\ni{\bf Structure 4} We first take differences of the position vectors $\underline{q}_I$ so as to cancel out any reference to $0$, 
thus forming the {\it relative Lagrange coordinates}
\be
\uir_{IJ} = \uiq_J - \uiq_I \mbox{ }  ;
\ee
see Fig \ref{Jac-Med-Ineq-Fig-2}.b).

\mbox{ }

\ni{\bf Remark 22} The $\ur_{IJ}$ are not all linearly independent; for instance 
\be
\uir_{13} = \uir_{12} + \uir_{23} \mbox{ } . 
\ee
One can get around this by picking a basis of two of the $\uir^{IJ}$ to work with, e.g.\ $\ur_{12}$ and $\ur_{13}$. 
Using such a basis moreover breaks the $\uq_I$ description's diagonality of moment of inertia, kinetic term and so on. 

\mbox{ }

\ni{\bf Structure 5} This diagonality can be restored, however, by considering not particles but particle clusters (subsystems): {\it relative Jacobi coordinates}, 
$\underline{R}_1$ and $\underline{R}_2$. 
These are widely used in e.g.\ Celestial Mechanics \cite{Marchal} and in Molecular Physics \cite{LR97}.    
Now not only relative point-or-particle separations are required but also separations between different clusters' centres of mass 
(indicated by crosses in Fig \ref{Jac-Med-Ineq-Fig-2}.c).

\mbox{ }

\ni{\bf Remark 23} Choosing $r_{12} = R_1$ is one of a possible 3 choices. 
We denote this by $\uR^{(3)}$ alias $\uR^{(c)}$, 
and the clusters with $\ur_{23}$ and $\ur_{31}$ as their first relative Jacobi coordinate 
               by $\uR^{(1)}$ alias $\uR^{(a)}$ 
			  and $\uR^{(2)}$ alias $\uR^{(b)}$ respectively. 

\mbox{ }

\ni{\bf Example 3}	The minimal $N = 3$ case has moreover 3 choices of clustering, 
corresponding to {\it labelling ambiguity} in picking a first relative separation $\uir_{IJ}$ in forming it. 
We refer to this as the {\it base pair} of the clustering and to the remaining point-or-particle $K$ as the {\it apex}.  
The diagonality condition then requires the other relative Jacobi vector to be from the centre of mass of two particles to the third particle.  
Thus, for equal masses -- the case considered in the current paper --
\beq
\uiR_1           =           \uiq_3 - \uiq_2                                  \mbox{ } \mbox{ } \mbox{ and } \mbox{ } \mbox{ } 
\uiR_2  \mbox{ } =  \mbox{ } \uiq_1 - \frac{\uiq_2 + \uiq_3}{2}                                                       \mbox{ } , 
\label{Jac-Defs}
\eeq
where, using the well-known elementary formula for reduced mass
\beq
\frac{1}{\mu} \mbox{ } =  \mbox{ } \frac{1}{m_1} \mbox{ } + \mbox{ } \frac{1}{m_2} \mbox{ } , 
\eeq 
\beq
\mu_1  \mbox{ } =  \mbox{ } \frac{1}{2}                               \mbox{ } \mbox{ and } \mbox{ } 
\mu_2  \mbox{ } =  \mbox{ } \frac{2}{3}                                                     \mbox{ } . 
\label{mu}
\eeq 
\ni{\bf Definition 11} Let us additionally term the 3-point-or-particle model in $d \geq 2$'s first and second relative Jacobi vectors the {\it base} and {\it median} vectors.

\mbox{ }

\ni{\bf Remark 24} This refers to the equal-mass case, $\mbox{CoM(base)}$ is at the centre of the base chosen by the clustering. 
The second Jacobi vector moreover runs from this to the opposing vertex.
Due to this, `median' is indeed an appropriate name for it in the $d \geq 2$ case's triangular configuration. 

\mbox{ }

\ni {\bf Remark 25} We have the following identifications.
\be 
R_{1}^{s_i} = s_i \mbox{ } , 
\ee 
\be
R_{2}^{s_i} = m_i \mbox{ } . 
\ee 
\ni{\bf Definition 12} We furthermore denote $\mu_1$ by $\mu_{\sss}$ -- {\it side Jacobi mass}    -- 
                                         and $\mu_2$ by $\mu_{\sm}$:   {\it median Jacobi mass}.  

\mbox{ } 

\ni{\bf Remark 26} Take note also of the ratio of the square roots of these masses in the equal-mass case under consideration in the current paper:   
\be
\sqrt{\frac{\mu_{\sm}}{\mu_{\sss}}} = \frac{2}{\sqrt{3} } \mbox{ } . 
\ee
This is moreover indeed the value of section 2's $\kappa$ constant, which is now identified conceptually and physically with this square-rooted mass ratio,  
\be
\kappa := \sqrt{\frac{\mu_{\sm}}{\mu_{\sss}}} = \frac{2}{\sqrt{3} } \mbox{ } . 
\ee 
\ni{\bf `Jacobian' Motivation 1} for placing medians on equal footing to sides.  

\mbox{ }

\ni{\bf Corollary 6} i) 
\be
R_2^{(i)\,2} \mbox{ } = \mbox{ } \kappa^{-2} \, J^{(ij)} R_1^{(i)\,2} \mbox{ } , 
\label{1}
\ee
inverting to ii) 
\be
R_1^{(i)\,2} \mbox{ } = \mbox{ } \kappa^2    \, J^{(ij)} R_2^{(i)\,2} \mbox{ } . 
\label{2}
\ee
{\underline{Proof}} Substitute (\ref{Jac-Defs}) into the Linear Algebra form of the sides--medians relation (\ref{0.8}). $\Box$

\section{Mass-weighted Jacobi coordinates}

\ni Let us next introduce mass-weighted Jacobi coordinates.  
On the one hand, these further simplify previous sections' workings. 
On the other hand, the Kendall and Hopf coordinates that are the subsequent focus of this paper are built out of mass-weighted Jacobi coordinates.  

\mbox{ }

\ni{\bf Structure 6} {\it Mass-weighted relative Jacobi coordinates} are given by   
\beq
\urho_a := \sqrt{\mu_a} \uR_{a} \mbox{ } , 
\label{rho-R}
\eeq 
where the $a$-index takes values 1 and 2.   

\mbox{ }

\ni{\bf Structure 7} {\it Mass-weighted relative Jacobi separations} are the magnitudes of the preceding, 
\beq
\rho_a := \sqrt{\mu_a}R_a 
\eeq 
\ni{\bf Remark 27} Thus computationally, 
\beq
\rho_1^{(a)} := \sqrt{\mu_1}R_1^{(a)}  \mbox{ } = \mbox{ } \frac{a}{\sqrt{2}} \mbox{ } \mbox{ and cycles } , 
\eeq 
alongside 
\beq
\rho_1^{(a)}          :=           \sqrt{\mu_1}R_1^{(a)} 
             \mbox{ }  = \mbox{ }  \frac{m_a}{\sqrt{2}} \mbox{ } \mbox{ and cycles } \mbox{ }
             \mbox{ }  = \mbox{ }  \frac{\sqrt{2 \, b^2 + 2 \, c^2 - a^2}}{  2  \sqrt{2}  } .   
\eeq
\ni{\bf Definition 13} The {\it first mass-weighted moment sums} are, for $a$ = 1, 2, 
\be 
F_a \mbox{ } := \mbox{ } \sum_i\rho_a^{(i)}                                        \mbox{ } .
\ee 
\ni{\bf Remark 28} The mass-weighted Jacobi separations are moreover related to the more widely used {\it partial moments of inertia} $I_a$ by 
\be 
I_a = \rho_a\mbox{}^2                                                              \mbox{ } .
\label{I-rho} 
\ee 
In particular, with clustering labels explicit,  
\be 
I_1^{(a)}          =          \rho_1^2 
                   =          \mu_1 R_1\mbox{}^2 
          \mbox{ } = \mbox{ } \frac{a^2}{2} \mbox{ } \mbox{ and cycles } , 
\ee 
and 
\be 
I_2^{(a)}          =           \rho_2^2 
                   =           \mu_2 R_2\mbox{}^2 
		  \mbox{ } = \mbox{ } \frac{m_a\mbox{}^2}{2} \mbox{ } \mbox{ and cycles } .  
\ee
\ni{\bf Definition 14} The {\it second mass-weighted moment sums} are, for $a$ = 1, 2, 
\be 
Z_a \mbox{ } := \mbox{ } \sum_i\rho_a^{(i)2} 
    \mbox{ }  = \mbox{ } \sum_i I_a^{(i)}                                          \mbox{ } . 
\ee 
\ni{\bf Definition 15} More familiarly, summing over disjoint partial moments rather than over clusters, the {\it total moment of inertia} is  
\be 
I^{(a)} := I_1^{(a)} + I_2^{(a)}                                                       \mbox{ } . 
\ee 
The definition here is that the total object is the sum of all disjoint partial contributions.  

\mbox{ }

\ni{\bf Lemma 8} (Democratic formula for the moment of inertia)
\be
I \mbox{ } =  \mbox{ } \frac{  a^2 + b^2 + c^2  }{  3  }   
  \mbox{ } =: \mbox{ } \left\langle side^2 \right\rangle                               \mbox{ } , 
\label{Iabc}
\ee
where $\left\langle \mbox{ } \right\rangle$ denotes `democratic average'.

\mbox{ }

\ni {\underline{Proof}} See \cite{2-Herons} for the first equality; the term `democratic average' is explained in Sec \ref{DI}.  

\mbox{ } 

\ni The second equality follows from the standard definition of average, applied here over the 3 possible clusterings. $\Box$  

\mbox{ }

\ni{\bf Remark 29} The right hand side of the first equality being cluster-independent, we rewrite 
\be 
I := I_1^{(a)} + I_2^{(a)} \mbox{  or cycles } . 
\ee 
\ni{\bf Remark 30} While trivial to prove, the Lemma's second equality none the less has {\it conceptual content} 
as regards giving a sharp interpretation for the computational right hand side of the first equality. 

\mbox{ }

\ni{\bf Definition 16} Let $\bigsigma_{\sR\sS\sS}$ denote the mass-weighted version of  $S_{\sR\sS\sS}$ and  
                          $\bigmu_{\sR\sS\sS}$    the mass-weighted version of  $M_{\sR\sS\sS}$. 

\mbox{ }						  
						 
\ni{\bf Corollary 7} 
\be 
I \mbox{ } =  \mbox{ } 2 \, \bigsigma_{\sQ}\mbox{}^2 
  \mbox{ } =  \mbox{ } 2 \, \mbox{\Large$\mu$}_{\sQ}\mbox{}^2                                                                   \mbox{ } ,
\ee 
or, inverting, 
\be 
\bigsigma_{\sQ}\mbox{}^2           = \mbox{\Large$\mu$}_{\sQ}\mbox{}^2  
              \mbox{ } =  \mbox{ } \frac{I}{2}                                                                                             \mbox{ } .
\ee
\ni{\bf Remark 31} This signifies that, in mass-weighted space, cluster summed or averaged second moments of side or of median are equivalent, 
and both are readily convertible to the moment of inertia up to a constant of proportionality.  

\mbox{ }  

\ni{\bf Corollary 8} i) 
\be
\rho_2^{(i)\,2} = J_{ij} \rho_1^{(i)\,2} \mbox{ } , 
\label{5}
\ee
inverting to ii) 
\be
\rho_1^{(i)\,2} = J_{ij} \rho_2^{(i)\,2} \mbox{ } . 
\label{6}
\ee
iii) A slightly tidier version in terms of partial moments of inertia is
\be
I_2^{(i)} = J^{(ij)} I_1^{(i)} \mbox{ } , 
\label{5b}
\ee
inverting to iv) 
\be
I_1^{(i)} = J_{ij} I_2^{(i)} \mbox{ } . 
\label{6b}
\ee
{\underline{Proof}} See \cite{2-Herons}. 

\mbox{ }

\ni{\bf Remark 32} Note that \cite{2-Herons} the mass-weighting cleans out the awkward numerical factor of $4/3 = \kappa^2$ from eqs (\ref{0.6}-\ref{0.8}). 

\mbox{ }

\ni{\bf Remark 33} Thus summing up over the clusterings gives 
\be
Z_1 \mbox{ } := \mbox{ } \sum_i \rho_1^{(i)\,2} 
    \mbox{ }  = \mbox{ } \sum_i I_1^{(i)} 
	\mbox{ } = \mbox{ }  \sum_i I_2^{(i)} 
	\mbox{ } = \mbox{ }  \sum_i \rho_2^{(i)\,2} =: Z_2 \mbox{ } , 
\label{7}
\ee
which is also now free of numerical factors [as compared to (\ref{0.10})'s factor of 3/4].  
This signifies that median partial moments of inertia contribute -- over all clusterings -- the {\sl same amount} as side moments of inertia.  
This amounts to a statement of {\it Jacobi-isotropy} among the sums over all clusters of the partial moments of inertia. 
Furthermore, dividing both sides by 3, 
\be
\left\langle \, \rho_1^{2} \, \right\rangle = \left\langle \, \rho_2^{2} \, \right\rangle \mbox{ }  
\label{7.b}
\ee
casts this interpretation in the form of equality between (the cluster-averaged median partial moment of inertia) 
                                                      and (the (cluster-averaged  side partial moment of inertia).

\section{Consequent Jacobi separation inequalities}

\ni Passing to mass weighted Jacobi versions of Sec 3's inequalities requires knowing how the quantities involved scale under mass-weighting.  

\mbox{ }

\ni{\bf Lemma 9} i)  In Fig \ref{6-variables}, everything in column 1 is a side-vector and median-scalar, 
                     whereas                   everything in column 2 is a median-vector and side-scalar.

\mbox{ } 					
					
\ni              ii) ${\cal S}$, ${\cal G}$, ${\cal M}$ and ${\cal H}$ are mass-weighting scalars.

\mbox{ }

\ni{\underline{Proof}} %
Using the conceptualization in the following footnote,\footnote{The language this is phrased in is a similarity tensor calculus (i.e. the constant scalefactor version of a conformal tensor calculus 
as is more familiarly used e.g. in the GR initial-value problem \cite{York72}.  
See \cite{TRiPoD, ABook} for further examples and theory of similarity tensor calculus including comparison with full conformal tensor calculus. 
Moreover, in the current paper's application, one has to distinguish between mass-weighting by median Jacobi masses and by side Jacobi masses. 
Due to this feature, the current paper's quantities are in general similarity {\it bitensors}; 
see e.g. DeWitt's \cite{DeWitt65} for exposition of the bitensor version of the usual tensor calculus on curved spaces.
Note moreover that objects scaling as $\mu_{\sss}^{\alpha}\mu_{\sm}^{\beta}$ and 
                                      $\mu_{\sss}^{\gamma}\mu_{\sm}^{\delta}$ 
							      for $\alpha + \beta = \gamma + \delta$ {\sl} can be added, in contradistinction with many other bitensor calculi.} 
this is clear from the simple side and median dependencies of these quantities. $\Box$

\mbox{ }

\ni{\bf Remark 34} In alotting primary ratios, I used a maximal amount of what would be mass-weighted scalars in the first place, for convenience in the working. 
This means that no mass-weighted versions of these (or notation for such) need be introduced at this stage.  

\mbox{ }

\ni{\bf Lemma 10} \ni i) Area is a mass-weighting side--median bivector,
\be
\bigupalpha\mbox{rea}         =          \sqrt{\mu_{\sss}\mu_{\sm}} \, Area 
                       \mbox{ } = \mbox{ } \frac{    Area   }{   \sqrt{3}   } \mbox{ } .
\ee
ii) ${\cal F}$, ${\cal E}$ and ${\cal I}$ are all median-vectors and side-covectors. 

\mbox{ }

\ni{\underline{Proof}} i) 
\be
\mbox{Area} \mbox{ } = \mbox{ } \frac{1}{2}\{\uR_1 \times \uR_2 \}_3 
            \mbox{ } = \mbox{ } \frac{1}{2\sqrt{\mu_1\mu_2}} \, \{\urho_1 \times \urho_2 \}_3 
			\mbox{ } = \mbox{ } \frac{\bigupalpha\mbox{rea}}{\sqrt{\mu_s\mu_m}}                  \mbox{ } . 
\ee
ii) The first two are linear in medians and reciprocally linear in the sides, whereas the last goes as $Area$/side$^2$ and thus scales as median/side as well.

\mbox{ }

\ni{\bf Remark 35} As set up, while working at the level of ratios, everything is a biscalar or a vector-covector, whose weight is then $\kappa$. 

\mbox{ }

\ni{\bf Definition 17} Denote the {\it mass-weighted medimeter per unit perimeter} by   
\be
\Psi     := \kappa \,  {\cal F} \mbox{ } ,  
\ee 
the {\it mass-weighted root sum of medians per unit perimeter} by  
\be
\Lambda  := \kappa \, {\cal L}  \mbox{ } ,  
\ee 
We do not use the {\it mass-weighted isoperimetric ratio} 
\be
\bigiota := \kappa \, {\cal I} \mbox{ } , 
\ee
so as to maintain the [0, 1] range. 
To be clear, in each case here, `mass weighted' means that both the numerator and the denominator are mass-weighted.  

\mbox{ }

\ni{\bf Theorem 3}   
\be
\kappa^{-1} \mbox{ } \leq \mbox{ } \Psi \mbox{ } \leq \mbox{ } \kappa
\label{mw-Rat-Ineq-1}
\ee 
\ni{\underline{Proof}} Mass-weight (\ref{Rat-Ineq-1}). $\Box$.  

\mbox{ }

\ni{\bf Remark 36} Mass-weighting has multiplicatively centred the inequalities' bounds, meaning that one bound is now the reciprocal of the other. 

\mbox{ }

\ni{\bf Theorem 4}  
\be
{\cal I} \mbox{ } \leq \mbox{ } \Psi 
         \mbox{ } \leq \mbox{ } \Lambda 
                    =           {\cal S}^{-1} 
		 \mbox{ } \leq \mbox{ } {\cal I}^{-1}{\cal G}^{-3} \mbox{ } .
\label{mw-rat-ineq}
\ee
\ni{\underline{Proof}} Mass-weight (\ref{Rat-Ineq-2}). $\Box$

\mbox{ }

\ni{\bf Remark 37} On the one hand, Theorem 1 places shape-independent bounds on the mass-weighted medimeter-to-perimeter ratio. 
On the other hand, Theorem 2 places shape-dependent bounds on this and the root sum median to perimeter ratio, 
the shape dependence of which is manifested as powers of the isoperimetric ratio and arithmetic-to-geometric mean side raito shape variables.
As regards lower bounds, for 
\be 
{\cal I} \mbox{ } < \mbox{ } \kappa^{-1} \mbox{ } , 
\ee 
Theorem 2's shape-dependent one is more stringent, whereas for 
\be 
{\cal I} \mbox{ } > \mbox{ } \kappa^{-1} \mbox{ } , 
\ee
Theorem 1's shape-independent one is. 
As regards upper bounds, for  
\be 
{\cal I} \, {\cal G}^3 \mbox{ } > \mbox{ } \kappa^{-1} \mbox{ } ,
\ee 
Theorem 2's shape-dependent one is more stringent, whereas for 
\be 
{\cal I} \, {\cal G}^3 \mbox{ } < \mbox{ } \kappa^{-1} \mbox{ } ,
\ee
Theorem 1's shape-independent one is. 
The corresponding cross-over's critical values, and consideration of the set of shapes entailed by each case, is part of \cite{A-Perimeter}.  

\mbox{ }

\ni{\bf Remark 38} The bounding significance of the equilateral triangle $E$ and uniform collinear shape $U$ is clearly inherited from the mass-unweighted precursor in Remark 19.

\section{Relational space and shape space coordinates in terms of mass-weighted Jacobi vectors}

\ni{\bf Structure 8} The mass-weighted Jacobi separations $\rho_1$, $\rho_2$, and 
\be
\Phi  = \mbox{arccos}\left(  \frac{  \urho_1 \cdot \urho_2  }{  \urho_1\urho_2  }    \right) \mbox{ } : \label{Swiss} 
\ee 
a `Swiss-army-knife' relative angle as per Fig \ref{Jac-Med-Ineq-Fig-2}.d) provide coordinates on relational space, corresponding to quotienting out translations and rotations.

\mbox{ }

\ni{\bf Structure 9} On the other hand, the {\it unit Jacobi vectors} $\underline{n}_i := \urho_i/\sqrt{\rho}$ give coordinates on the preshape 3-sphere,  
\be
\rho_{1x}\mbox{}^2 + \rho_{1y}\mbox{}^2 + \rho_{2x}\mbox{}^2 + \rho_{2y}\mbox{}^2 \mbox{ } = \mbox{ } \rho^2   \mbox{ } ,
\ee
or 
\be
n_{1x}\mbox{}^2 + n_{1y}\mbox{}^2 + n_{2x}\mbox{}^2 + n_{2y}\mbox{}^2 \mbox{ } = \mbox{ } 1   \mbox{ } .
\ee
Preshape space are in general much simpler to provide metrics for than relational spaces or the below shape spaces.
Because of this, the simplest route to determining the metric on a shape space is via the preshape space and not the relational space. 

\mbox{ }

\ni{\bf Structure 10} Quotienting out all of translations, dilations and rotations, one arrives at the shape space.  
For triangles, two incipient coordinates for this are $\Phi$ and the (mass-weighted relative Jacobi separation) ratio 
\be
{\cal R} := \frac{\rho_2}{\rho_1}  \mbox{ } . 
\ee
In terms of this, the shape space metric is 
\be 
\d s^2 = 4 \, \frac{\d {\cal R}^2 + {\cal R}^2\d\Phi^2}{  (  1 + {\cal R}^2  )^2  }
\ee
Then the well-known substitution 
\be
{\cal R} = \mbox{tan} \frac{\Theta}{2} \mbox{ } 
\ee 
yields the obvious form of the standard spherical metric
\be 
\d s^2 = \d {\Theta}^2 + \mbox{sin}^2 \Theta \, \d\Phi^2 \mbox{ } . 
\ee
In this geometrical context, moreover, ${\cal R}$ is now recognized as the stereographical radius.  

\mbox{ }

\ni{\bf Structure 11} The relational space metric is then recovered from this by the standard topological and metric coning construct.  

\mbox{ }

\ni{\bf Remark 39} While this section's results enter much of the shape theory of the triangle, 
we note that $S$, $M$, $S_{\sG}$ and $M_{\sG}$ are {\sl complicated} expressions in terms of $\Theta$ and $\Phi$. 
$I$ on the other hand, is simple, but has nothing else simple to form a ratio with apart from $Area$.  
Moreover, $\bigupalpha rea$ per unit moment of inertia was already analyzed in \cite{III}.

\section{The Hopf map and 2/3 of its shape-theoretic interpretation}

\ni There is however a further formulation related to the previous Section's which does interact more directly with the current paper's $S$, $M$, $S_{\sG}$ and $M_{\sG}$ objects.
\ni This is the Hopf formulation.

\mbox{ }

\ni{\bf Structure 12} The Hopf map is from the 3-sphere to the 2-sphere. 
As already outlined in Paper I, this is realized by the reduction from the preshape 3-sphere to the shape sphere.
Indeed, this provides a derivation that the triangleland shape space is a sphere.

\mbox{ }

\ni{\bf Remark 40} In 1-$d$ it is immediately clear how to represent an $\{n - 1\}$-sphere $\FrS(N, 1) = \FrP(N, 1)$ within ${\cal R}(N, 1) = \Frr(N, 1) = \mathbb{R}^n$.
For 3 particles in $d \geq 2$, however, this takes a more complicated form. 
While there is a flat relative space $\Frr(3, 2)$, this is $\mathbb{R}^4$ rather than $\mathbb{R}^3$.  
So how does one sit a 2-sphere in a flat 4-space in an equable manner? 
(I.e.\ making equal use of each component of $\urho_1$ and $\urho_2$).  
The trick is to concatenate the Hopf map with the obvious codimension-1 embeddings of $\mathbb{S}^2$ and $\mathbb{S}^3$: 
\beq
\mathbb{R}^4  \mbox{ } \stackrel{    \mbox{\scriptsize unit sphere map}    }{    \longrightarrow    }                 \mbox{ }
\mathbb{S}^3  \mbox{ } \stackrel{    \mbox{\scriptsize Hopf fibration (I.159)}    }{    \longrightarrow    } \mbox{ } 
\mathbb{S}^2  \mbox{ } \stackrel{    \mbox{\scriptsize embedding map}    }{    \longrightarrow    }                   \mbox{ }
\mathbb{R}^3                                                                                                          \mbox{ } . 
\eeq 
\ni In Shape Theory, this sequence of spaces realizes, respectively, relative space, preshape space, shape space and relational space.
\ni We refer to the overall map in this equation as the `extended Hopf map'.  
This indeed provides a set of Cartesian directions within an ambient $\mathbb{R}^3$, 
which are built up democratically from the obvious incipient Cartesian directions of $\mathbb{R}^4$. 

\mbox{ }

\ni{\bf Structure 13} In this shape-theoretic setting, and with respect to a choice of clustering to define the relative Jacobi coordinates 
underpinning the $\underline{\rho}_i$, $\slTheta$ and $\slPhi$, the extended Hopf map takes the form 
\beq
Hopf_x :=  I \, \mbox{sin}\,\slTheta\,\mbox{cos}\,\slPhi 
        =  2 \, \rho_1 \rho_2 \, \mbox{cos}\,\slPhi 
	    =  2 \{\underline{\rho}_1 \cr \underline{\rho}_2\}_3 \mbox{ } ,
\label{Hopf-x}
\eeq
\beq
Hopf_y :=  I \, \mbox{sin}\,\slTheta\,\mbox{sin}\,\slPhi 
        =  2 \, \rho_1 \rho_2\,\mbox{sin}\,\slPhi 
	    =  2 \, \underline{\rho}_1 \cdot \underline{\rho}_2 \mbox{ } ,
\label{Hopf-y}
\eeq
\beq
Hopf_z =  I \, \mbox{cos}\,\slTheta 
       =  \rho_2\mbox{}^2 - \rho_1\mbox{}^2   	                                   \mbox{ } . 
\label{Hopf-z}
\eeq
\ni Here each equation's first equality just recasts the Hopf Cartesian coordinates in standard spherical polar coordinates,  
with the moment of inertia $I = \rho^2$ playing the role of radius.     
Each equation's second equality amounts to the shape-theoretic relations between the $\rho_i$ and $I$, $\slTheta$ and $\slPhi$.  
The third equality, when present, just incorporates standard basic formulae for dot and cross products.

\mbox{ }

\ni{\bf Structure 14} There is moreover a unit Cartesian direction version of this obtained by dividing each term by $I$.
Using unit relative Jacobi coordinates, this gives 
\beq
hopf_x :=  \mbox{sin} \, \slTheta\,\mbox{cos}\,\slPhi 
        =  2 \, n_1 n_2 \ ,\mbox{cos} \, \slPhi 
	    =  2 \{\underline{n}_1 \cr \underline{n}_2\}_3                   \mbox{ } ,
\label{hopf-1}
\eeq
\beq
hopf_y := \mbox{sin} \, \slTheta\,\mbox{sin}\,\slPhi 
        =  2 \, n_1 n_2\,\mbox{sin} \, \slPhi 
	    =  2 \, \underline{n}_1 \cdot \underline{n}_2                    \mbox{ } ,
\label{hopf-2}
\eeq
\beq
hopf_z :=  \mbox{cos} \, \slTheta 
        =  n_2\mbox{}^2 - n_1\mbox{}^2                                   \mbox{ } .   
\label{hopf-3}
\eeq
These can be readily checked to obey the on-2-sphere condition
\be
hopf_x\mbox{}^2 + hopf_y\mbox{}^2 + hopf_z\mbox{}^2 = 1                   \mbox{ } . 
\label{hopf-on-sphere}   
\ee
\ni This subsection is well-known and applied in Molecular Physics \cite{LR97}, but has received little attention in Shape Statistics.

\mbox{ }

\ni{\bf Remark 41} Clearly $hopf_y$ is tetra-area per unit moment of inertia, 
measuring departure from collinearity, and a further place where specifically tetra-area is geometrically realized.  
In \cite{+Tri, FileR, III}, we show moreover that $hopf_x$ measures departure from isoscelesness, by which we term it anisoscelesness, $aniso$.  
                                         and that $hopf_z$ measures departure from equality of the side-base and median partial moments of inertia. 
We term this quantity ellipticity, $ellip$. 										 

\mbox{ }

\ni{\bf `Hopfian motivation'} for median coprimality is firstly that a property treating sides and medians on an equal footing 
is itself placed on the same footing as the more traditionally considered departures from isosceles and collinear configurations. 
Secondly, as shown in \cite{2-Herons}, the Hopf quantities retain their form up to sign upon exchanging sides and medians 
(and these signs are an allowed freedom in choosing Cartesian axes).

\section{Ellipticity, regularity, tallness and flatness}

\ni{\bf Definition 18} A triangle is {\it regular} if its base and median partial moments of inertia are equal:  
\be 
I_{s}    = I_{m} \mbox{ } \Leftrightarrow \mbox{ } 
{\rho_2} = {\rho_1}  \mbox{ } .  
\ee 
\ni{\bf Remark 43} There are moreover 3 notions of regularity, corresponding to the 3 clustering choices, 
\be 
I_{s}^{(i)} = I_{m}^{(i)} \mbox{ } . 
\ee
\ni{\bf Remark 44} Equilateral triangles are indeed regular, giving the following further Jacobi-separation-uniformity characterization.
A triangle is equilateral if base--median equality holds in two of its Jacobi clusterings.
[This condition clearly holds in {\sl all three} of its clusterings, but it holding in two {\sl implies} it holds in the third as well.] 

\mbox{ }

\ni{\bf Remark 45} To see this (and for later use), note that in space itself, due to intervention of the Jacobi masses, the regularity condition translates to 
\be
\frac{    R_2    }{    R_1    } \mbox{ } = \mbox{ } 
%
%
\frac{    \sqrt{3}    }{    2    }                              \mbox{ } , 
\label{Regular}
\ee  
which is then immediately recognizable as the equilateral triangle's base-to-height ratio.  
Regularity can thus conversely be formulated in entirely flat space geometrical terms 
-- without any reference to mass weighting or moments of inertia -- as `the base and median are in the proportion found in the equilateral triangle'.  

\mbox{ }

\ni {\bf Definition 19} A triangle is {\it tall} -- denoted $\mT$ -- if the inequality 
\be 
   I_{s} \mbox{ } < \mbox{ } I_{m} \mbox{ }  \Leftrightarrow \mbox{ } 
{\rho_2} \mbox{ } < \mbox{ } {\rho_1} 
\ee 
holds. 
It is {\it flat} -- denoted $\mF$ -- if the inequality 
\be
I_{s} \mbox{ } > \mbox{ } I_{m} \mbox{ }  \Leftrightarrow \mbox{ } {\rho_2} \mbox{ } > \mbox{ } {\rho_1} 
\ee
holds instead.

\section{Democracy invariants}\label{DI}

\ni{\bf Definition 20} The existence of multiple choices of clustering furthemore motivates considering linear transformations between different clusterings,  
termed `{\it democracy transformations}' in the Molecular Physics literature \cite{Zick, ACG86, LR95}.  

\mbox{ }
																		   
\ni{\bf Structure 8} Given a quantity in one clustering, one can cycle over all three clusterings and add up (or average) to make a democratic version of the quantity in question.  
This usually works.  

\mbox{ }
																		   
\ni{\bf Example 1} Ellipticity, for instance, is a cluster-dependent tallness quantity.
Its cluster independent version is shown below not to work, but there is a linear rather than quadratic counterpart which does work.

\section{Democratic ellipticity quantifier and inequalities satisfied by it}

\ni{\bf Definition 21} The democratic version of ellipticity is 
\be 
\left\langle Ellip \right\rangle \mbox{ } := \mbox{ } \bigmu_{\sQ}\mbox{}^2 - \bigsigma_{\sQ}\mbox{}^2          \mbox{ } ,  
\ee 
whereas the corresponding pure-ratio quantity is, normalizing by 
\be
Z_{\sT}         :=           \bigmu_{\sQ}\mbox{}^2 + \bigsigma_{\sQ}   
              \mbox{ } =  \mbox{ } \frac{1}{3} \, \sum_i \left(    \rho_1^{(i) \, 2} + \rho_2^{(i) \, 2}    \right)    \mbox{ } ,  
\ee 
\be
\left\langle ellip \right\rangle  \mbox{ } :=  \mbox{ } \frac{\left\langle Ellip \right\rangle}{Z_{\sT} } \mbox{ } .  
\ee
\ni{\bf Remark 46} However, since $Z_1 = Z_2$ by (\ref{7}), $\left\langle Ellip \right\rangle$ and thus $\left\langle ellip \right\rangle$ are zero.
Thus democratic ellipticity in the immediate senses of the preceding Sec is a sterile concept. 

\mbox{ }

\ni{\bf Remark 47} This suggests seeking an alternative ellipiticity quantifier whose democratic form remains shape-theoretically nontrivial. 
The obvious candidate for a such follows from considering the linear counterpart of the above construct, as follows. 

\mbox{ }

\ni{\bf Definition 22} The {\it linear ellipticity} is given by    
\be
Lin\mbox{--}Ellip                 := \rho_2 - \rho_1 \mbox{ } .  
\ee 
\ni{\bf Remark 48} $Lin$-$Ellip$'s zero correponds to
\be 
2 \, a^2 = b^2 + c^2 \mbox{ } .
\ee
Thus equilaterality $a = b = c$ solves, but not uniquely so; given $b$, $c$ we can always pick 
\be 
a \mbox{ } = \mbox{ } \sqrt{\frac{b^2 + c^2}{2}}
\ee 
to have a linearly-regular shape.  

\mbox{ }

\ni{\bf Remark 49} The normalized version, $lin$-$ellip$, turns out to have no extrema, 
its {\it extremal values} occurring at the ends of its allowed range: -- 1 for the binary coincidence-or-collision shape $B$ and + 1 for the uniform collinear configuration, $U$.  

\mbox{ }

\ni{\bf Structure 9} The democratic version of these are, firstly,  
\be 
\left\langle Lin\mbox{--}Ellip \right\rangle \mbox{ } :=  \mbox{ } \bigmu_{\sA} - \bigsigma_{\sA} \mbox{ } . 
\ee 
Secondly, normalizing by 
\be
F_{\sT}                     :=          \bigmu_{\sA} + \bigsigma_{\sA} 
                       \mbox{ } = \mbox{ } \frac{1}{3} \sum_i \left(    \rho_1^{(i)} + \rho_1^{(i)}    \right)                                      \mbox{ } ,  
\ee 
where `T' stands for total average: {\sl all} Jacobi coordinates, corresponding to all mass-weighted sides {\sl and} all mass-weighted medians,
\be
\left\langle lin\mbox{--}ellip \right\rangle \mbox{ } :=  \mbox{ } \frac{\left\langle Lin\mbox{--}Ellip \right\rangle}{F_{\sT} } 
                                             \mbox{ }  =  \mbox{ } \frac{\bigmu - \bigsigma}{\bigmu + \bigsigma} 
								             \mbox{ }  =  \mbox{ } \frac{\Psi - 1}{\Psi + 1}                                                        \mbox{ } . 
\ee
\ni{\bf Remark 50} To determine the zeros of $\left\langle lin\mbox{--}ellip \right\rangle$, a systematic if pedestrian method is as follows. 

\mbox{ }
 
\ni 1) Isolate two roots on one side. Square to leave a total of two roots in contention. 

\mbox{ }

\ni 2) Isolate these on the same side of the ensuing equation, and square again. 
This leaves one root in contention.
 
\mbox{ }

\ni 3) Finally isolate this root on one side and square again.

\mbox{ }

\ni 4) An `Olympian amount' of cancellations occur in steps 2) and 3) 
(we leave it to the reader to see if a more elegant working can be found that avoids/explains all these cancellations) leaving us with
\be
S^2 \left( 27 \, S_{\sG}\mbox{}^3 - 64 \, M_{\sG}\mbox{}^3 \right) = 0 \mbox{ } .
\ee
Thus, since $S = 0$ is the maximal collision $O$ which is excluded by non-normalizability, we arrive at the condition  
\be
{\cal G} \mbox{ } := \mbox{ } \kappa^{-1}                              \mbox{ } . 
\ee
\ni{\bf Remark 51} Now, for sure, the equilateral triangle obeys this, but there are clearly other values which do as well. 
\cite{A-Perimeter} analyzes the subsequent cubic equation in $a^2$ given $b^2$ and $c^2$. 
{\sl positive} solutions to this cubic support $\left\langle lin\mbox{--}ellip \right\rangle$, so there can be 0 to 3 solutions given specific 2-side data $b, c$.  
See \cite{A-Perimeter} for an extremal analysis of $\left\langle lin\mbox{--}ellip \right\rangle$ as well.

\mbox{ }

\ni{\bf Corollary 9}   
\be
- \frac{2 - \sqrt{3}}{2 + \sqrt{3}} \mbox{ } \leq  \mbox{ } \left\langle lin\mbox{--}ellip \right\rangle  
                                    \mbox{ } \leq  \mbox{ }       \frac{2 - \sqrt{3}}{2 + \sqrt{3}}                                                     \mbox{ } .
\ee
\ni{\bf Remark 52} This is symmetrically placed, as befits a two-tailed quantity. 
This result also prompts a renormalized re-issuing of the definition of linear ellipticity as follows, so that the range now be the standard 2-tailed one from -- 1 to +1.

\mbox{ }

\be
\left\langle \widetilde{lin\mbox{--}ellip} \right\rangle \mbox{ } := \mbox{ } \frac{2 + \sqrt{3}}{2 - \sqrt{3}} \, \frac{\bigmu - \bigsigma}{\bigmu + \bigsigma} 
								                         \mbox{ }  = \mbox{ } \frac{2 + \sqrt{3}}{2 - \sqrt{3}} \, \frac{\Psi - 1}{\Psi + 1}            \mbox{ } . 
\ee
\ni{\bf Remark 53} For the equilateral triangle $E$, $\left\langle \widetilde{lin\mbox{--}ellip} \right\rangle = 0$. 
$\left\langle \widetilde{lin\mbox{--}ellip} \right\rangle_{\sm\sa\sx}$ is at the binary coincidence-or-collision $B$, whereas  
$\left\langle \widetilde{lin\mbox{--}ellip} \right\rangle_{\sm\si\sn}$ is at the uniform collinear configuration $U$.  
Among the splinters, these are (Fig \ref{S(3, 2)-Splinters}.b) respectively, the extreme spear-head and the extreme pickaxe-heads. 

\mbox{ }

\ni{\bf Theorem 5}   
\be
- 1 \mbox{ } \leq \mbox{ } \left\langle \widetilde{lin\mbox{--}ellip} \right\rangle 
    \mbox{ } \leq \mbox{ } 1                                                                                                                            \mbox{ } .
\ee
\ni{\bf Theorem 6} 
\be
\frac{{\cal I} - 1}{{\cal I} + 1}                                                             \mbox{ } \leq \mbox{ }  
\frac{2 - \sqrt{3}}{2 + \sqrt{3}} \left\langle \widetilde{lin\mbox{--}ellip} \right\rangle    \mbox{ } \leq \mbox{ } 
\frac{1 - {\cal I} \, {\cal G}^3}{1 + {\cal I} \, {\cal G}^3}                                                                                                                 \mbox{ }   .
\ee
\ni{\bf End-Remark 1} A subsequent step is to plot and optimize ${\cal S}$, ${\cal M}$, ${\cal G}$, ${\cal H}$, $\Psi$, $\Lambda$ 
and $\left\langle \widetilde{lin\mbox{--}ellip} \right\rangle$ over the shape sphere, much as \cite{III} did for mass-weighted area per unit moment of inertia.  
This is considered in the subsequent paper \cite{A-Perimeter}, alongside plotting further geometrically significant functions for triangles over the shape sphere of triangles. 

\mbox{ }

\ni{\bf End-Remark 2} Another subequent investigation is of quadrilateral and polygon inequalities for Jacobi and generalized--Hopf (e.g. Gell-Mann $SU(3)$ \cite{QuadI}) quantities. 

\mbox{ }

\ni{\bf Acknowledgments} I thank Chris Isham and Don Page for previous discussions, 
                                 Jimmy York Jr. for providing great motivation in my youth as reflected in part in footnote 1,  and 
                                 Malcolm MacCallum, Reza Tavakol, Jeremy Butterfield and Enrique Alvarez for support with my career. 
This paper is dedicated to those who think differently as regards what they need in life.

\vspace{10in}


\end{document}